\documentclass[lettersize,journal]{IEEEtran} 
\usepackage{amsmath,amsfonts}
\usepackage{algorithmic}
\usepackage{algorithm}
\usepackage{array}
\usepackage[caption=false,font=footnotesize,labelfont=rm,textfont=rm]{subfig}
\usepackage{textcomp}
\usepackage{stfloats}
\usepackage{url}
\usepackage{makecell}
\usepackage{verbatim}
\usepackage{multirow}
\usepackage{graphicx}
\usepackage{cite}
\usepackage{xcolor}
\allowdisplaybreaks
\begin{document}

\title{QoE-Driven Multi-Task Offloading for Semantic-Aware Edge Computing Systems}

\author{Xuyang Chen,~\IEEEmembership{Student Member,~IEEE}, Daquan Feng,~\IEEEmembership{Member,~IEEE}, Wei Jiang,~\IEEEmembership{Member,~IEEE}, \\Qu Luo,~\IEEEmembership{Member,~IEEE}, Gaojie Chen,~\IEEEmembership{Senior Member,~IEEE}, and Yao Sun,~\IEEEmembership{Senior Member,~IEEE}
\thanks{This work was supported in part by the National Science Fund for Excellent Young Scholars under Grant 62422112, in part by the National Key Research and Development Program of China under Grant 2024YFB2808903, in part by the Guangdong Key Areas Research and Development Program under Grant 2022B0101010001, and in part by the Shenzhen Key Research and Development Program under Grant KJZD20230923114707016, and in part by the Fundamental and Interdisciplinary Disciplines Breakthrough Plan of the Ministry of Education of China (JYB2025XDXM406), and in part by the Fundamental Research Funds for the Central Universities, Sun Yat-sen University, under Grant No.24hytd010. (Corresponding author: Daquan Feng)}
\thanks{Xuyang Chen and Daquan Feng are with the College of Electronics and Information Engineering, Shenzhen University, Shenzhen 518060, China (email: chenxuyang2021@email.szu.edu.cn; fdquan@szu.edu.cn).}
\thanks{Wei Jiang is with the Institute of Cyberspace Security, Zhejiang University of Technology, Hangzhou 310023, China (e-mail: weijiang@zjut.edu.cn).}
\thanks{Qu Luo is with the 5GIC \& 6GIC, Institute for Communication Systems, University of Surrey, GU2 7XH Guildford, U.K. (e-mail: : q.u.luo@surrey.ac.uk; gaojie.chen@surrey.ac.uk).}
\thanks{Gaojie Chen is with the School of Flexible Electronics (SoFE)
\& State Key Laboratory of Optoelectronic Materials and Technologies, Sun
Yat-sen University, Guangdong, 510220, China. (e-mail: gaojie.chen@ieee.org)}
\thanks{Yao Sun is with the James Watt School of Engineering, University of Glasgow, Glasgow, G12 8QQ, U.K. (email: Yao.Sun@glasgow.ac.uk).}
}



\maketitle

\begin{abstract}
Mobile edge computing (MEC) provides low-latency offloading solutions for computationally intensive tasks, effectively improving the computing efficiency and battery life of mobile devices. However, for data-intensive tasks or scenarios with limited uplink bandwidth, network congestion might occur due to massive simultaneous offloading nodes, increasing transmission latency, and affecting task performance. 
In this paper, we propose a semantic-aware multi-task offloading framework to address the challenges posed by limited uplink bandwidth. By introducing a semantic extraction factor, we balance the tradeoff among transmission latency, computation energy consumption, and task performance. To measure the offloading performance of different tasks, we design a unified and fair quality of experience (QoE) metric that includes execution latency, energy consumption, and task performance. We formulate the QoE optimization problem as a Markov decision process (MDP) and exploit the semantic-aware multi-agent proximal policy optimization (MAPPO) reinforcement learning algorithm to jointly optimize the semantic extraction factor and communication and computing resources allocation to maximize overall QoE. Experimental results show that the proposed method achieves a 12.68\% improvement in user QoE compared with the semantic-unaware approach. Moreover, the proposed approach can be easily extended to models with different user preferences.
%
\end{abstract}


\section{Introduction}
\label{intro}
\IEEEPARstart{T}{he} sixth generation (6G) of mobile communication is demonstrating its revolutionary potential, portending the arrival of an era of ultra-massive connectivity and the Internet of Things (IoT) \cite{ZHANG202260}. The volume of data generated by a vast number of devices is experiencing an explosive surge. Various compute-intensive services, such as holographic communication, extended reality (XR), and autonomous driving continue to evolve, which increasingly burden user equipments (UEs) with heavy computational demands. However, the limited computational power and battery capacity of UEs make it challenging to support a multitude of computationally intensive tasks. 

Cloud computing is considered capable of alleviating the computational pressure on UEs to some extent \cite{MCC-TCC}. Offloading tasks to the cloud, which has abundant computing resources, significantly reduces local computational pressure. However, the high latency associated with cloud offloading is intolerable for latency-sensitive tasks. Furthermore, the centralized model of cloud computing poses data privacy and security issues. To address the aforementioned challenges, mobile edge computing (MEC) is considered an effective solution. Edge servers (ES) near the user can provide considerable computing and storage resources, which can be fully utilized by UEs distributed. Compared to cloud computing, MEC significantly reduces latency, while its distributed computing resources ensure user privacy and security.

Nevertheless, when tasks involve substantial data transfers, such as high-definition (HD) video processing \cite{Zhu-TCSVT-2020}, virtual reality (VR) applications \cite{wiservr,Cle-CM-2020}, or real-time data analytics \cite{RN1965,Zhao-TMC-2023}, the increased volume of data imposes higher demands on channel quality to meet latency requirements. In such cases, semantic communication (SemCom) can provide an effective solution \cite{RN2999}. SemCom, different from the traditional bit workflow based on Shannon's information theory \cite{Weaver-1953}, has demonstrated great potential for enhancing the efficiency and reliability of data transmission across various task scenarios, including text/audio \cite{RN1824,RN1740,RN1739,RN2954}, image \cite{RN1778,RN1735,RN1891}, video \cite{LeonardoMM2020,RN1755,RN1752}, XR \cite{RN1754,wiservr,RN1807}, and multi-modal \cite{RN1817,RN1907,RN1700}. By transmitting a compact semantic representation of task-relevant data, substantial bandwidth can be saved, ensuring high quality of service (QoS).

Given these considerations, it is essential to design a semantic-aware MEC system by combining MEC with SemCom \cite{RN1902}. By performing semantic extraction on the data to be offloaded, UEs can filter out data irrelevant to the task, thereby reducing network traffic. In \cite{yan2023qoe}, the authors defined a semantic entropy to quantify the semantic information of different tasks. Deep Q-Network (DQN) and many-to-one matching algorithms are used to solve the joint optimization problem of semantic compression, channel allocation, and power distribution in multi-cell multi-task scenarios. A semantic-aware resource allocation method was proposed in \cite{RN1880}, introducing a semantic-aware factor to characterize the relationship between semantic compression rate and computational overhead, aiming to balance the degree of semantic compression. The authors of \cite{ji2024computational} proposed a semantic-aware cloud-edge-end collaborative networking method to utilize distributed computing resources and also proposed a multi-task offloading system in \cite{ji2023resource}, where a unified QoE metric was designed.

However, these works lack consideration for task execution performance and comprehensive QoE design. 6G strives to provide stable, reliable, and high-quality personalized services to each user \cite{science-china}. Preferences often vary among different users and types of services. For instance, in MEC networks, users dealing with latency-sensitive tasks may prefer lower latency, those pursuing task performance may favor better task execution accuracy, while users with limited battery power might opt for task offloading strategies that consume less power. Therefore, a good QoE metric should encompass the preferences of different users. Additionally, current research lacks comprea hensive consideration of the inherent heterogeneity across multi-modal tasks. Since each user may need to handle different tasks, the complexity and evaluation metrics of different tasks vary. For example, bimodal tasks like visual question answering (VQA) typically require more processing time and computational resources compared to text translation, and their evaluation metrics are entirely different.


Based on the above considerations, we design a resource allocation scheme for semantic-aware MEC networks that supports multi-modal tasks. We propose a semantic-aware factor to balance computation, communication, and task execution performance. To the best of our knowledge, this work is the first semantic-aware task offloading framework that jointly incorporates execution latency, execution energy consumption, and task performance into the QoE formulation. Additionally, a fair QoE metric is introduced to evaluate the utility of multi-modal task offloading, with support for user preference settings. The main contributions of this paper are summarized as follows.
\begin{itemize}
    \item We propose a novel semantic-aware task offloading system for multi-tasks. The SemCom systems for image classification, text classification, and VQA are first constructed to compress source data into compact semantic representations. The semantic extraction factor is used to characterize the degree of data compression, which affects the task's execution latency, energy consumption, and execution performance.
    \item We design a unified and fair QoE evaluation metric that addresses the challenges posed by the diverse distribution of user task types and the lack of uniform task evaluation criteria. Specifically, the QoE metric comprises three components: task execution latency, task execution energy consumption, and task execution performance. We employ a unified logistic function to normalize these three components, thereby obtaining a unified and comparable QoE.
    \item The joint optimization for semantic extraction factor, computational resources, and communication resources is performed using the semantic-aware MAPPO algorithm, maximizing the overall QoE. Users solve for the optimal channel allocation, power allocation, semantic extraction level, and computation offloading scheme according to their observed task sizes and channel conditions.
    \item Comprehensive experiments demonstrate the effectiveness of the proposed approach. Compared to the semantic-unaware method, the proposed algorithm demonstrates a 12.68\% improvement in user QoE. Additionally, the proposed approach can meet user preferences for various metrics such as execution latency, energy consumption, and execution performance.
\end{itemize}

The remainder of this paper is organized as follows. Section \ref{related_works} introduces the existing literature and related works. Section \ref{sys} describes the semantic-aware transmission and computation model, and defines the unified QoE metric, and formulates the problem as maximizing overall QoE. Section \ref{method} introduces solving the QoE maximization problem using the MAPPO algorithm, providing training details. Section \ref{simulation} presents relevant experimental details and simulation results. Finally, Section \ref{conclusion} outlines our conclusions.

\section{Related Works}
\label{related_works}
MEC networks have attracted widespread attention from academia and industry \cite{MEC-survey} due to its ability to enhance the computing capabilities of UEs and improve the QoS. Numerous studies have adopted numerical optimization techniques to derive optimal or near-optimal solutions \cite{MEC-Deng,RN1916,RN1918,RN2976,RN2977,RN2983}. In \cite{MEC-Deng}, the authors optimized the allocation of communication, computation, and energy resources based on a perturbed Lyapunov algorithm, maximizing system throughput under the stability constraints of task and energy queues. A power minimization method with task buffer stability constraints was proposed in \cite{RN1916}, where the Lyapunov optimization algorithm was used to solve computing offloading strategies. The requirements for ultra-reliable and ultra-low latency (URLLC) tasks was considered in \cite{RN1918}, applying extreme value theory to minimize user power consumption while satisfying probabilistic queue length constraints. In \cite{RN2976}, a multi-branch deep neural network (DNN) model was developed to minimize energy consumption in device–edge collaborative inference, and the trade-off between branch selection and inference accuracy was analytically quantified.

However, the aforementioned model-based approaches are typically computationally intensive, further burdening user devices. Moreover, model-based methods often require multiple iterations to find the optimal offloading strategy, which is intolerable for latency-sensitive tasks and impractical in rapidly changing wireless channel conditions. To address these issues, resource allocation and computation offloading methods based on Deep Reinforcement Learning (DRL) were proposed \cite{RN1919,RN2980,RN3006,RN2991,RN1920,RN1921}. The authors of \cite{RN1919} modeled the computation offloading as a multi-stage stochastic mixed-integer nonlinear programming (MINLP) problem and solved it using a combination of Lyapunov optimization and a DRL model, integrating the advantages of both model-based and DRL approaches. In \cite{RN2980}, two offloading schemes based on soft actor critic (SAC), i.e., centralized SAC offloading (CSACO) and distributed SAC offloading (DSACO) were proposed for routing and scheduling between adjacent ESs to alleviate congestion in the core network. CSACO collects global states and makes decisions to approximate the global optimum, while DSACO enables each ES to make decisions based on local states, supporting decentralized learning and reducing communication overhead. In \cite{RN3006}, the authors designed a multi-agent deep deterministic policy gradient (MADDPG)-based path planning scheme to minimize the total energy consumption of unmanned aerial vehicle (UAV) systems. During training, the agents exploit global information, whereas in actual deployment, they make decisions relying solely on local observations. In \cite{RN2991}, a distributed DRL algorithm was proposed to minimize task offloading delay. The authors enhanced the long-term value estimation by integrating long short-term memory (LSTM) and double deep Q-network (DDQN). 

The optimization objectives in task offloading can be either single or multiple. Single-objective optimization may aim to minimize offloading delay, energy consumption, or operational cost \cite{RN2994, RN2995, RN3007,RN2978, RN2982, RN2984, RN2985}. In \cite{RN2994}, the authors jointly optimized server deployment and user offloading decisions in a wireless backhaul edge network, proposing an efficient genetic algorithm-based approach to minimize the average service delay of users. An aerial hierarchical MEC system was introduced in \cite{RN2995}, which supports dynamic task offloading across three tiers of nodes: local devices, unmanned aerial vehicles (UAVs), and high-altitude platforms (HAPs). In \cite{RN3007}, the authors constructed a distributed many-to-many matching model that allows task owners and edge nodes to negotiate offloading and payment decisions through multiple rounds of interactive and autonomous bargaining, thereby preserving user privacy. In \cite{RN2978}, a novel intelligent framework integrating edge computing with few-shot learning was developed to enable semantic inference based on limited data samples.

Multi-objective optimization in task offloading aims to simultaneously consider multiple performance goals. Typically, several objectives are combined through weighted summation or by treating some as constraints \cite{RN2988, RN2989, RN2992, RN2993, RN3008, RN2981, RN2987,10999053}. In \cite{RN2988}, a parameterized DQN-based method was proposed to solve the online service placement and resource allocation problem. Two DNNs are employed to estimate the optimal service placement decisions and the corresponding computing resource allocation, respectively. The authors of \cite{RN2989} developed a collaborative inference model that dynamically selects the partition point of deep learning models. By designing an experience-sharing mechanism, neighboring devices exchange Q-values to accelerate the optimization of the policy network. In \cite{RN2992}, a three-tier MEC network architecture supporting end–edge–cloud collaborative caching was proposed, aiming to jointly optimize content retrieval delay and handover latency. The authors of \cite{RN2993} were the first to model a task offloading problem with spatio-temporal load awareness, effectively preventing the overload of certain satellites while avoiding resource underuse in others.

In MEC networks, task characteristics such as heterogeneity, dependency, and priority significantly influence offloading decisions \cite{RN3027, RN3025, RN2990, RN3024, RN3022, RN3023, RN2986}. In \cite{RN3027}, the authors designed a meta-learning mechanism that enables the algorithm to rapidly adapt to dynamic changes in the edge–cloud environment. In \cite{RN3025}, the authors investigated the multi-task offloading problem in vehicular edge computing (VEC), considering dynamic and flexible task topologies, heterogeneous resource demands, and variations in communication and computation capabilities. The authors of \cite{RN3022} modeled interdependent tasks as a directed acyclic graph (DAG) and proposed a priority-based scheduling method derived from task cost, addressing collaborative offloading under multi-task dependencies in satellite IoT systems. In \cite{RN2986}, the authors jointly optimized task partitioning ratios and user association strategies in a multi-user, multi-edge-node MEC scenario, supporting the offloading of both independent subtasks and sequentially dependent subtasks.

Semantic encoding leverages powerful feature extractors to drastically reduce required bandwidth and has demonstrated strong potential in text/audio  \cite{RN1824,RN1740,RN1739,RN2954}, image/video \cite{RN1778,RN1735,RN1891,LeonardoMM2020,RN1755,RN1752}, 3D/XR \cite{RN1754,wiservr,RN1807}, and multi-modal transmission \cite{RN2601,RN2595}. \cite{RN1824} is one of the earliest implementations of text-based semantic encoding. The authors designed an end-to-end deep learning semantic communication system that jointly optimizes semantic and channel encoding, effectively extracting semantic information from text and enhancing robustness against channel noise. Subsequently, in \cite{RN1824}, the authors further proposed a lightweight distributed semantic communication system that significantly reduces model size and complexity through weight pruning and quantization, enabling deployment on resource-constrained IoT devices. In \cite{RN1739}, the authors developed a knowledge graph (KG)–based text semantic communication system, in which textual data are represented as ``entity–relation–entity" triples. A robust voice semantic transmission scheme was proposed in \cite{RN2954}, where the authors adopt a hierarchical design that separates audio compression from digital encoding and transmission, while fully exploiting the contextual prediction capabilities of large language models (LLMs).

Semantic transmission technologies have been widely applied in the field of visual data. In terms of semantic image transmission, a deep learning–based joint source–channel coding (JSCC) method was proposed in \cite{RN1778}, , which maps image pixel values to complex-valued channel input symbols, thereby enabling the end-to-end wireless image transmission system. In \cite{RN2975}, the authors developed a wireless image semantic communication framework that jointly optimizes system capacity, pixel-level distortion, semantic-level distortion, and perceptual quality. The information bottleneck theory was employed to balance the trade-off between channel bandwidth ratio and transmission robustness. A content-aware image semantic transmission scheme was presented in \cite{CHEN20251205}, where images are divided into regions of interest (ROI) and regions of non-interest (RONI) according to downstream tasks. By adjusting quantization levels, the algorithm significantly reduces the amount of encoded data. In \cite{11207608}, a multi-level generative image semantic transmission approach was proposed, in which both textual and visual features are used to guide image denoising to generate realistic images.

In terms of semantic video transmission, a semantic transmission scheme for video conferencing was proposed in \cite{RN1755}, which significantly reduces data redundancy by transmitting only the keypoint changes between adjacent frames. In \cite{11175641}, a latent diffusion model (LDM)–based semantic-aware adaptive bitrate streaming framework was designed. By combining LDM with FFmpeg technology, the framework optimizes bandwidth utilization and storage efficiency for real-time video streaming. An ultra-low bitrate semantic video transmission scheme was developed in \cite{RN2692}, where only the first frame image and subsequent audio signals are transmitted to drive the lip movements of individuals in the video. A video transmission scheme supporting semantic self-correcting and fine-grained bitrate control is proposed in \cite{chen2025qoe}, where video frames are encoded into a semantic codebook space and matched with the current bandwidth through active packet loss.

Due to the massive volume of XR and 3D data, many researchers have explored semantic encoding algorithms for data transmission. In \cite{wiservr}, a semantic transmission framework tailored for VR content was developed. The concept of semantic location graph (SLG) was introduced, which significantly reduced data transfer volume by separately transmitting the semantic features of static objects and dynamic objects. In \cite{RN2929}, neural radiance fields (NeRF) were combined with nonlinear transform coding for the first time to develop an end-to-end semantic transmission system for free-viewpoint 3D scenes. An interest-aware communication framework for point cloud video (PCV) streaming was proposed in \cite{RN2765}. The authors designed a two-stage ROI selection mechanism that dynamically filters key content based on user motion trajectories and saliency analysis, substantially reducing network and computational resource consumption while maintaining video quality. In \cite{AItransfer}, an end-to-end deep learning network integrating feature extraction and reconstruction was designed. By transmitting only the key features of point clouds instead of raw data, this approach markedly reduced the bandwidth requirements for real-time point cloud video streaming.

\begin{table*}[t]
\centering
\caption{Comparison Between Related Works and Ours}
\begin{tabular}{|c|c|c|c|c|c|c|c|c|}
\hline
References & \makecell{Semantically \\ Adjustable} & Multi-Task & \makecell{Energy \\ Efficiency} & \makecell{Communication \\ Efficiency} & \makecell{Execution \\ Performance} & \makecell{Resource \\ Allocation} & DRL Algorithm & \makecell{Hybrid \\ Action Space}\\ \hline
\cite{RN3015} & & & & \checkmark & \checkmark & \checkmark & & \\ \hline
\cite{RN3016} & \checkmark & & \checkmark & \checkmark & \checkmark & \checkmark & & \\ \hline
\cite{RN3017} & \checkmark & & \checkmark& & & \checkmark& & \\ \hline
\cite{RN3018}& \checkmark& & & \checkmark& & & \checkmark& \\ \hline
\cite{RN3020}& & & & \checkmark& & \checkmark& \checkmark& \\ \hline
\cite{RN3014}& \checkmark& & \checkmark& & & \checkmark& & \\ \hline
\cite{RN3012}& & & \checkmark& \checkmark& & \checkmark& & \\ \hline
\cite{RN3011}& \checkmark& & & \checkmark& & \checkmark& & \\ \hline
\cite{RN3010}& \checkmark& & \checkmark& \checkmark& & \checkmark& \checkmark& \\ \hline
\cite{RN3013}& \checkmark& & \checkmark& & & \checkmark& & \\ \hline
\cite{RN3004}& & & \checkmark& \checkmark& & \checkmark& & \\ \hline
\cite{yan2023qoe}& &\checkmark& & \checkmark& \checkmark& \checkmark& \checkmark& \\ \hline
\cite{RN2699}& \checkmark& & \checkmark& \checkmark& & \checkmark& & \\ \hline
\cite{ji2023resource}& & \checkmark & \checkmark & \checkmark & & \checkmark & \checkmark & \checkmark \\ \hline
\cite{ji2024computational}& & & \checkmark & \checkmark & & \checkmark & \checkmark & \checkmark \\ \hline
our work & \checkmark & \checkmark & \checkmark & \checkmark & \checkmark & \checkmark & \checkmark & \checkmark \\ \hline
\end{tabular}
\label{related_works_table}
\end{table*}

The aforementioned semantic transmission schemes are all based on single-modality. However, in current multi-user and multi-scene environments, data from multiple modalities are often generated. Many studies have therefore sought to exploit multi-modal data to develop more efficient compression methods or design transmission frameworks that support multiple modalities. In \cite{Jiang_AAAI}, a multi-modal image compression framework was proposed, which for the first time utilized the semantic information from textual descriptions as prior knowledge to guide image compression, thereby improving reconstruction quality at extremely low bitrates. In \cite{RN1907}, the first unified end-to-end semantic communication system supporting multi-modal data was developed, capable of handling diverse modality tasks within a single model. In \cite{RN1700} a Transformer-based architecture for multi-user semantic communication was introduced, which efficiently extracts semantic information for different tasks, addressing the challenges of semantic extraction in multi-modal and multi-user scenarios. In \cite{RN3003}, an intelligent agent system based on a large edge AI model (LAM) was designed for the first time, capable of processing multi-modal data in a unified manner, understanding user intent through natural language, and generating personalized wireless communication strategies.

Semantic encoding technology can further reduce uplink traffic in MEC networks, and many studies have integrated the two approaches \cite{RN3015, RN3016, RN3017, RN3018, RN3020}. In \cite{RN3004}, the authors utilized a shared probabilistic graph for semantic information compression and derived the lower bound of signal estimation error under imperfect CSI. The authors of \cite{RN3013} focused on DNN inference tasks and quantitatively analyzed the nonlinear relationship between task compression ratio and local computation ratio. A semantic-aware MEC network for sustainable next-generation consumer electronic devices was proposed in \cite{RN3010}. By extracting and transmitting the semantic information of tasks using semantic communication, the approach significantly reduces transmission latency and energy consumption. In \cite{RN3011}, a multi-base-station, multi-terminal MEC system was designed, which reduces computational overhead by dynamically adjusting the depth of neural network computations. The authors of \cite{RN3012} combined semantic communication with low-Earth orbit (LEO) satellite-based edge clouds, extending the semantic encoder deployment to both ground stations and satellites. In \cite{RN3014}, a universal task offloading framework was proposed, integrating semantic communication, non-orthogonal multiple access (NOMA), MEC, and user mobility awareness. This framework introduces adjustable semantic factors for semantic extraction of task data, enabling dynamic trade-offs between local computational overhead and data transmission volume.

Table \ref{related_works_table} presents a comparison between our work and existing works across several key dimensions, including semantic adaptation, multi-task support, energy efficiency, communication efficiency, task execution performance, resource allocation, DRL algorithms, and hybrid action spaces. Leveraging semantic extraction factors, this paper comprehensively considers latency, energy consumption, and task performance while supporting the offloading of diverse heterogeneous tasks in MEC networks.

\section{System Model and Problem Formulation}
\label{sys}
\begin{figure}[t]
\centering
\includegraphics[width=0.48\textwidth]{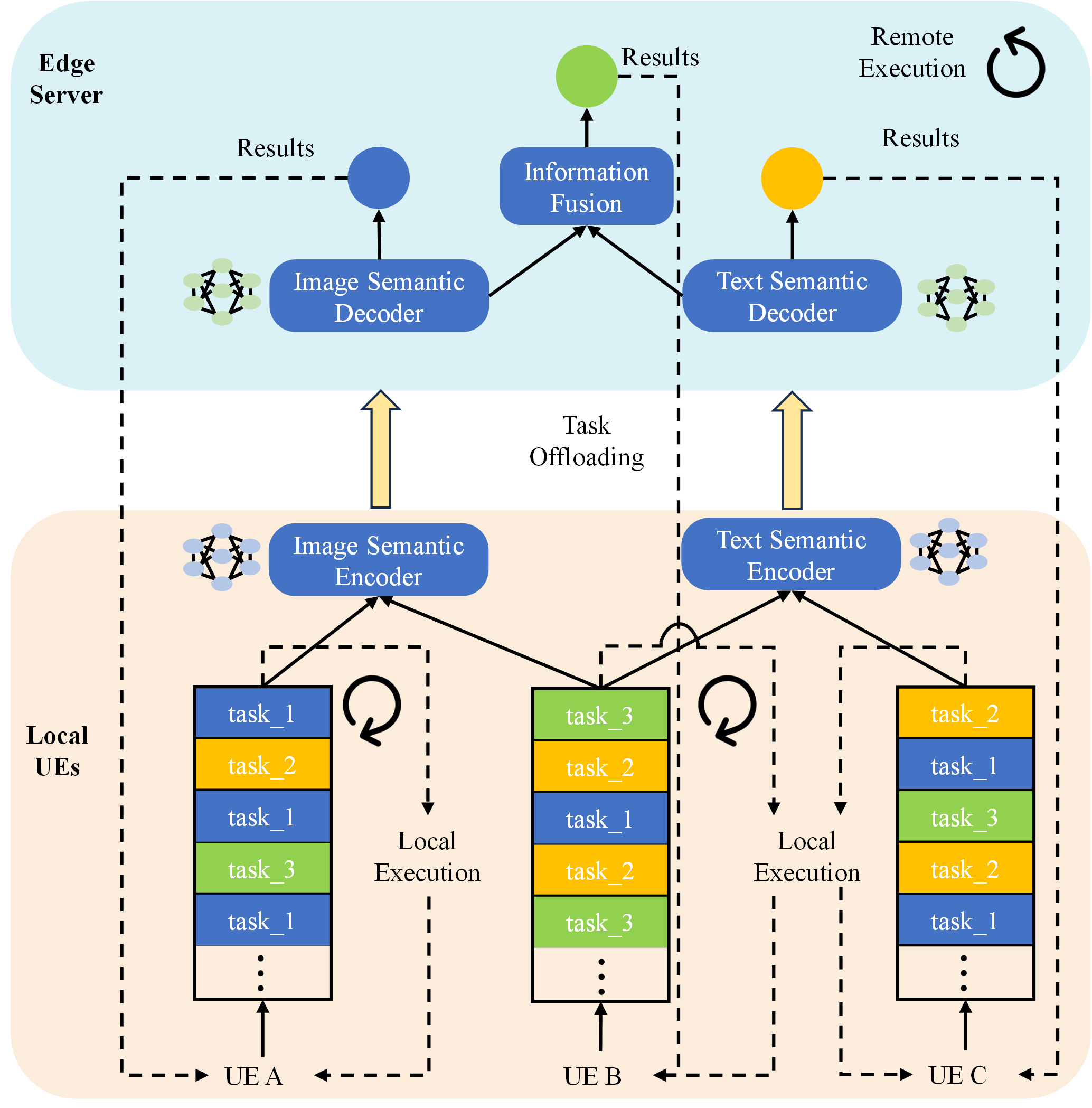}
\captionsetup{justification=raggedright}
\caption{Semantic-aware multi-task offloading system model. Each task is fed into a semantic codec specifically designed for its type.}
\label{sys-model}
\end{figure}

We propose a novel semantic-aware multi-task offloading system, as shown in Fig.~\ref{sys-model}, where UEs employ an orthogonal frequency-division multiple access (OFDMA) scheme. At the access point (AP), $K$ resource blocks are allocated to serve $N$ users at different sub-channels. The set of UEs and sub-channels are defined as $\mathcal{N} = \{1,2,\ldots,N\}$ and $\mathcal{K} = \{1,2,\ldots,K\}$. The length of the UE$_n$'s task queue is denoted as $Q_n$. The task currently being executed by the UE$_n$  is represented by $q^{\mathcal{M}}_n$, where $\mathcal{M}\in\{task_1, task_2, \ldots, task_m\}$. Hence, each UE may contain up to $m$ different types of tasks. We assume that tasks are atomic \cite{ji2023resource}, i.e., all tasks are either executed locally or offloaded to the ES. Considering the computational burden of semantic extraction and deep reinforcement learning, Graphic Processing Unit (GPU)-based architectures are employed to accelerate both training and inference in our proposed framework. The illustrations of the main notations are summarized in Table \ref{notations}.
\subsection{Semantic-Aware Model}
\begin{figure*}[t]
\centering
\includegraphics[width=0.9\textwidth]{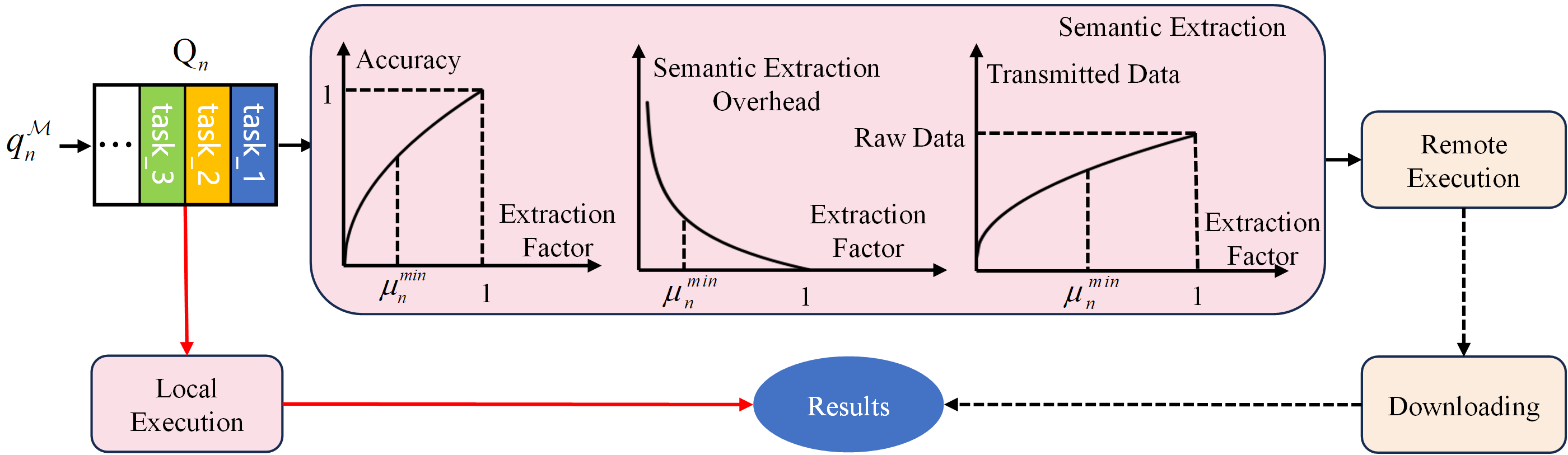}
\captionsetup{justification=raggedright}
\caption{Semantic extraction workflow. The semantic extraction factor alters the degree of semantic compression, affecting the computational burden of semantic extraction, the volume of data transmission, and the accuracy of task execution.}
\label{sem-model}
\end{figure*}

Traditional MEC offloading methods usually require transmitting all data to the ES. SemCom enables the extraction of task-relevant semantic information, reducing bandwidth consumption while enhancing robustness to channel variations. Based on this, we propose a semantic-aware multi-task offloading framework. Fig.~\ref{sem-model} illustrates the process of semantic-aware offloading. While semantic extraction indeed reduces the volume of data transmitted, the associated increase in semantic extraction latency and computational burden must also be comprehensively considered. Hence, we define a semantic extraction factor $\mu_n \in \left[\mu_n^{min}, 1\right]$.
The semantic extraction module determines the level of data compression. When $\mu_n=1$, the UE apply traditional offloading scheme instead of semantic-aware offloading. 

The proposed adaptive semantic extraction is inspired by the split computing paradigm in edge intelligence \cite{SPINN, SPLITEE}, where a deep learning model is partitioned between the UE and the ES. The semantic extraction module executes part of the model locally at the UE, while the remaining computation is handled at the ES using the transmitted intermediate semantic representations. The semantic extraction factor $\mu_n$ determines the balance of computational workload between the UE and ES, as well as the size of the transmitted features. While deeper local execution increases computation cost but reduces communication load, excessive compression of intermediate representations may degrade task performance at the server.
    
To maintain analytical tractability while capturing the essential trade-offs, without loss of generality, we model the semantic extraction overhead, transmission delay, and task performance as monotonic power-law functions of the semantic extraction factor. Such phenomenological modeling has been widely adopted in edge intelligence and communication systems to approximate complex deep learning behaviors \cite{11006481, 10287956,10032275}. It effectively captures the diminishing returns and non-linear sensitivity of computation and performance with respect to semantic compression. This abstraction enables joint optimization of semantic-aware offloading decisions without loss of generality, while avoiding the prohibitive complexity of explicit neural network-level modeling.

Transmitting all data can achieve better task performance, but it imposes a heavy transmission burden. The smaller the $\mu_n$, the greater the level of semantic compression, which significantly reduces the transmission load. However, semantic extraction introduces additional computational overhead and may decrease task performance on the ES. The minimum value $\mu_n^{min}$ is typically determined by task accuracy requirements and user battery capacity. Therefore, selecting $\mu_n$ requires a comprehensive evaluation of the current task computational load, channel environment, user battery capacity, and other relevant information.

\subsection{Transmission Model}
The channel gain from UE$_n$ to ES via sub-channel $k$ is denoted by $|g_{k,n}|^2 = |h_{k,n}|^2/d_n^{-\alpha}$, where $h_n \sim \mathcal{C} \mathcal{N}(0,1)$ is Rayleigh fading coefficient, and $d_n$ is the distance from UE$_n$ to ES, and $\alpha$ is the path loss exponent, and $\mathcal{C}\mathcal{N}(0,1)$ is the circularly-symmetric complex Gaussian distribution with zero mean and variance one. The maximum achievable offloading data rate from UE$_n$ to ES can be represented as
\begin{equation}
    R_{n} = \sum_{k=1}^KB\text{log}_2\left(1+\frac{x_{k,n}\left|g_{k,n}\right|^2p_n}{\sigma_z^2}\right),
\end{equation}
where $B$ is the sub-channel bandwidth, and $p_n$ denotes the transmit power from UE$_n$ to ES, and $\sigma_z^2$ is the additive white Gaussian noise (AWGN) noise power. $x_{k,n}$ denotes the sub-channel allocation indicators. $x_{k,n}=1$ indicates UE$_n$ is assigned to sub-channel $k$, $x_{k,n}=0$ otherwise. Therefore, the transmission latency can be expressed as
\begin{equation}
    t_n = \frac{\rho_ns_n\left(\mu_n\right)^p}{R_{n}},
\end{equation}
where $s_n$ is the normalized transmission data from UE$_n$ to ES, and $p>0$ is a constant parameter related to the specific task. $\rho_n$ is an offload indicator, where $\rho_n=1$ represents offloading tasks to ES, and $\rho_n=0$ denotes local computation. Compared to the amount of data required for task offloading, the computation results typically involve a much smaller data size \cite{KoTWC2018, Mach-survey2017}, therefore, we only consider the upload rate and latency.

\begin{table}[t]
\centering
\caption{Notations and Symbols}
\begin{tabular}{|>{\raggedright}m{0.1\textwidth}|p{0.3\textwidth}|}
\hline
\textbf{Notations} & \textbf{Explanations} \\
\hline
$\mathcal{N}$ & The set of all UEs \\ \hline
$\mathcal{K}$ & The set of all sub-channels \\ \hline
$\mathcal{M}$ & The set of task types \\ \hline
$\mu_n$ & The semantic extraction factor \\ \hline
$Q_n$ & The task length of UE$_n$ \\ \hline
$q_n^{\mathcal{M}}$ & The task currently being executed \\ \hline
$g_{k,n}$ & The channel gain from UE$_n$ to ES via sub-channel $k$ \\ \hline
$R_{k,n}$ & The maximum offloading rate from UE$_n$ to ES via sub-channel $k$ \\ \hline
$B$ & The channel bandwidth \\ \hline
$\rho_n$ & The offloading decision \\ \hline
$s_n$ & The transmitted message \\ \hline
$t_n$ & The transmission latency \\ \hline
$t_n^U$ & The computation time of UE$_n$ \\ \hline
$l_n^U$ & The GPU consumption executed by UE$_n$ \\ \hline
$l_n^{SE}$ & The basic GPU consumption of semantic extraction \\ \hline
$c_n$ & The GPU computation capability of UE$_n$ \\ \hline
$t^S$ & The computation time of ES \\ \hline
$l_n^S$ & The GPU consumption executed by ES \\ \hline
$c_s$ & The GPU computation capability of ES \\ \hline
$t_n^{\mathcal{M}}$ & The execution time of task $q_n^{\mathcal{M}}$ \\ \hline
$E_n^U$ & The energy consumption of UE$_n$ \\ \hline
$\kappa^U$ & The energy coefficient of UE$_n$ \\ \hline
$f_n$ & The GPU clock frequency of UE$_n$ \\ \hline
$E^S$ & The energy consumption of ES \\ \hline
$\kappa^S$ & The energy coefficient of ES \\ \hline
$f_s$ & The GPU clock frequency of ES \\ \hline
$\varepsilon_n^\mathcal{M}$ & The accuracy of task $\mathcal{M}$ \\ \hline
$\mathrm{QoE}_n$ & The QoE of UE$_n$ \\ \hline
$G_\mathcal{M}^t, G_\mathcal{M}^e, G_\mathcal{M}^a$ & The functions of execution time, energy consumption, and task accuracy score, respectively \\ \hline
$\omega_t, \omega_e, \omega_a$ & The weights of execution time, energy consumption, and task accuracy, respectively \\ \hline
$\lambda, \beta, \eta$ & The slope of the logistic function\\ \hline
\end{tabular}
\label{notations}
\end{table}

\subsection{Computation Model}
Each task can be executed in one of three modes: locally on the device, directly offloaded to the ES ($\mu_n=1$), or semantically pre-processed locally and then offloaded ($\mu_n^{min}\leq\mu_n<1$). The latency incurred by semantic extraction at the user should be considered in the overall task execution time. Therefore, the computation latency of UE$_n$ can be denoted as
\begin{equation}
\begin{aligned}
    t_n^U =
    \begin{cases}
        \frac{(1-\rho_n)l_n^U}{c_n},\ \text{if} \ \mu_n =1,\\
        \frac{(1-\rho_n)l_n^U}{c_n} + \frac{\rho_nl_n^{SE}}{\left(\mu_n\right)^qc_n},\ \mu_n^{min}\leq\mu_n<1, 
    \end{cases}
\end{aligned}
\end{equation}
where $q>0$ is a constant parameter relevant to the specific task, $l_n^U$ is the GPU consumption of the current task computed by UE$_n$, and $l_n^{SE}$ is the basic GPU consumption of semantic extraction. $l_n^U$ and $l_n^{SE}$ are related to the corresponding task $\mathcal{M}$. $k>0$ is a constant parameter. $c_n = \alpha_{ma}N_uf_n$ is the GPU computation capability of UE$_n$. $\alpha_{ma}=2$ corresponds to the floating-point operations (FLOPs) per cycle per core (FLOPs/cycle/core), where each instruction executes a fused multiply-add (FMA) operation. $N_u$ represents the number of compute unified device architecture (CUDA) cores available on UE$_n$. $f_n$ denotes the GPU clock frequency of UE$_n$. When tasks are offloaded to ES, the computation time of tasks on the ES can be expressed as
\begin{equation}
    t^S = \frac{\sum_{n=1}^Nl_n^S}{c_s},
\end{equation}
where $c_s=N_sf_s\alpha_{ma}$ is the GPU computation capability of ES. $l_n^S$ is the GPU consumption of the task offloaded by the UE$_n$ to the ES. $N_s$ denotes the number of CUDA cores available on ES. $f_s$ denotes the GPU clock frequency of ES. ES allocates its computational resources among users fairly, proportional to the amount of computation offloaded by each user. Combining the aforementioned factors, the processing latency of UE$_n$'s current task $q_n^\mathcal{M}$ can be expressed as
\begin{equation}
\begin{aligned}
    t^\mathcal{M}_n =& \rho_n(t_n+t_n^U+t^S) + (1-\rho_n)t_n^U\\
    =& \rho_n(t_n+t^S) + t_n^U.
\end{aligned}
\end{equation}

\subsection{Energy Consumption Model}
The energy consumption of UE$_n$ can be written as
\begin{equation}
    E_n^U = \kappa^Ut_n^U\left(f_n\right)^3 + p_nt_n,
\end{equation}
where $\kappa^U$ is the energy coefficient of UE$_n$. The energy consumption of ES can be described as \cite{ji2023resource}
\begin{equation}
    E^S=\kappa^St^S\left(f_s\right)^3,
\end{equation}
where $\kappa^S$ is the energy coefficient of ES. Therefore, the energy consumption of current task $q_n^\mathcal{M}$ can be expressed as
\begin{equation}
    E_n^\mathcal{M} = E_n^U+E^S.
\end{equation}

\subsection{Multi-Task Offloading Design and Problem Formulation}
The execution performance of all tasks in the task queue is represented as
\begin{equation}
    \varepsilon_n^\mathcal{M} = f^\mathcal{M}\left(\mu_n, \nu_n\right),
\end{equation} where $\varepsilon_n^\mathcal{M}$ is the accuracy of task $\mathcal{M}$, and $f^\mathcal{M}(\cdot)$ is the pretrained semantic transmission model for task $\mathcal{M}$, and $\nu_n$ is SNR. For a given fixed SNR, the task performance is modeled as a power-law function of $\mu_n$. Once the model training is completed, we can directly calculate the number of FLOPs required by the UE's semantic encoder to determine the basic GPU consumption $l_n^{SE}$ for semantic extraction. Similarly, we compute the number of FLOPs required by the ES's semantic decoder to determine the GPU consumption $l_n^S$ for the ES. To evaluate the performance of offloading tasks for different UEs, it is essential to define a unified QoE standard. 

\begin{figure}[t]
\centering
\includegraphics[width=0.48\textwidth]{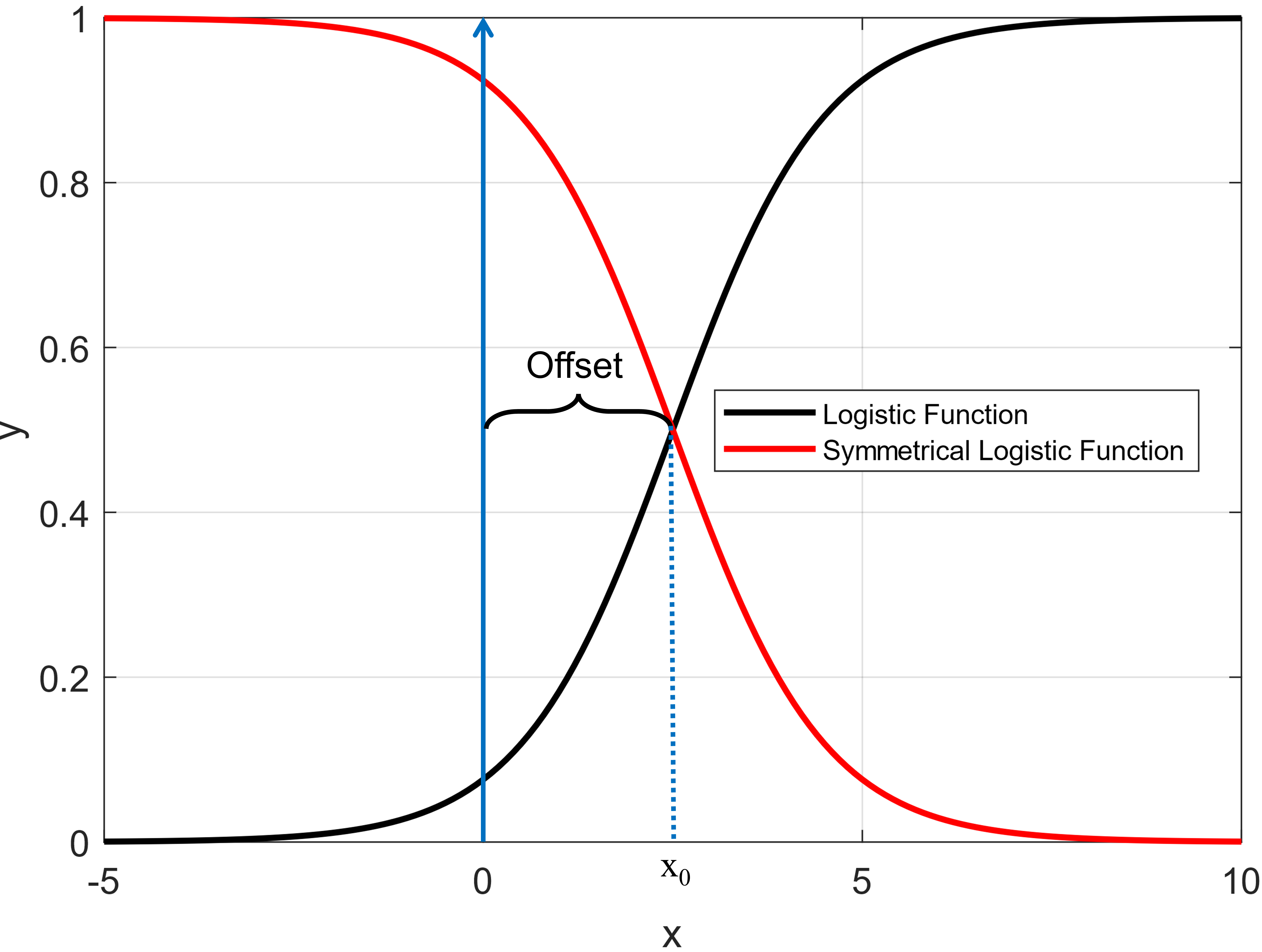}
\captionsetup{justification=raggedright}
\caption{QoE function. The logistic function serves as the scoring function for task accuracy, with its symmetric function scoring execution latency and energy consumption. The offset $x_0$ represents the latency, energy consumption, or task accuracy of local execution.}
\label{logistic}
\end{figure}

The performance of task execution is associated with computation time, energy consumption, and task accuracy. However, defining a fair and reasonable QoE is challenging. When tasks of different types are unevenly distributed among different UEs, directly calculating the weighted sum of energy consumption, computation time, and task accuracy is unfair. Typically, text tasks have lower computation time and energy consumption than image tasks, while image tasks have lower computation time and energy consumption than VQA tasks. Additionally, the accuracy of different tasks is usually incomparable. Inspired by \cite{Li-TCCN-2022, Hemmati-MM-2017, yan2023qoe}, we propose a QoE calculation method that can fairly compare task execution performance across different UEs:
\begin{equation}
\label{QoE}
    \begin{aligned}
        \mathrm{QoE}_n &= \frac{1}{Q_n}\sum_{q_n^\mathcal{M}=1}^{Q_n} (\omega_t G^t_{\mathcal{M}} + \omega_e G^e_{\mathcal{M}} + \omega_a G^a_{\mathcal{M}}) \\
                       &= \sum_{q_n^\mathcal{M}=1}^{Q_n} \left( \frac{\omega_t}{1+e^{-\lambda(t^l_n-t_n^\mathcal{M})}} + \frac{\omega_e}{1+e^{-\beta(E^l_n-E_n^\mathcal{M})}} \right. \\ 
                       & \left. + \frac{\omega_a}{1+e^{-\eta(\varepsilon_n^\mathcal{M}-\varepsilon^l_n)}} \right), 
    \end{aligned}
\end{equation}
where $\omega_t$, $\omega_e$, and $\omega_a$ are the weights of computation time, energy consumption, and task accuracy, respectively. The sum of $\omega_t$, $\omega_e$ and $\omega_a$ are equal to 1. $Q_n$ is the task length of UE$_n$.
$t^l_n=l_n^U/c_n$ denotes the task execution delay by UE$_n$. As shown in Fig.~\ref{logistic}, we use the logistic/sigmoid function and its symmetrical counterpart for QoE calculations. Corresponding latency score $G^t_{\mathcal{M}}$ and energy consumption score $G^e_{\mathcal{M}}$ are calculated using the symmetric logistic function, while accuracy score $G^a_{\mathcal{M}}$ is measured with the logistic function. $\lambda, \beta, \eta$ are the slope of the logistic function. We chose the logistic function for normalizing because it maps raw performance measures into a normalized interval $(0, 1)$, which is suitable for probability-like interpretations and ensures stability when combining terms. Moreover, the S-shaped curve of the logistic function effectively captures the diminishing marginal utility of QoE. For example, when network resources are already abundant, further increases in resources lead to only marginal improvements in QoE. The offset $x_0$ represents local execution latency, energy consumption, or accuracy. Specifically, $E_n^l=\kappa^Ut_n^l(f_n)^3$ denotes the task execution energy consumption by UE$_n$. $\varepsilon_n^l$ denotes the task execution accuracy by UE$_n$. That is, we normalize the QoE using the latency, energy consumption, and accuracy of tasks executed locally by the UE.

Mathematically, the optimization problem can be formulated as
\begin{subequations}
\label{optimization_problem}
\begin{align}
\textbf{(P0)} \quad  &\underset{\{\boldsymbol{\rho}, \boldsymbol{p}, \boldsymbol{f}, \boldsymbol{x},\boldsymbol{\mu}\}}{\operatorname{maximize}} \quad 
\sum_{n=1}^N\mathrm{QoE}_{n} \\
\label{opti-rho}
\text{s.t.}\quad & \rho_n \in \{0,1\}, \quad \forall n \in \mathcal{N},\\
\label{opti-x}
& x_{k,n} \in \{0,1\},  \quad \forall n\in \mathcal{N}, \forall k \in \mathcal{K},\\
\label{opti-k}
& \sum_{k=1}^{\mathcal{K}}x_{k,n} \leq \rho_n, \quad \forall n \in \mathcal{N},\\
\label{opti-n}
& \sum_{n=1}^{\mathcal{N}}x_{k,n} \leq 1, \quad \forall k \in \mathcal{K},\\
& \mu_n^{min} \leq \mu_n \leq 1, \quad \forall n \in \mathcal{N}, \\ \label{opti-pmax}
& p_n \leq p_{\max}, \quad \forall n \in \mathcal{N},\\ \label{opti-fmax}
& f_n \leq f_{\max}, \quad \forall n \in \mathcal{N},\\ \label{opti-tmax}
& t_n^\mathcal{M} \leq t_{\max}, \quad \forall n \in \mathcal{N},\\ \label{opti-emax}
& E_n^\mathcal{M} \leq E_n^{\max}, \quad \forall n \in \mathcal{N},\\ \label{opti-amax}
& \varepsilon_n^{\mathcal{M}} \geq \varepsilon_{\min}, \quad \forall n \in \mathcal{N},
\end{align}
\end{subequations}
where $\{\boldsymbol{\rho},\boldsymbol{p},\boldsymbol{f},\boldsymbol{x}, \boldsymbol{\mu}\}=\{\rho_n, p_n, f_n, x_{k,n}, \mu_n\}, \forall{n} \in \mathcal{N}, \forall{k} \in \mathcal{K}$. $\boldsymbol{x} \in \mathbb{R}^{K\times N}$ denote the collection of all sub-channel allocation indicators $x_{k,n}$. $x_{k,n}=1$ indicates UE$_n$ is assigned to sub-channel $k$, $x_{k,n} = 0$ otherwise. In (\textbf{P0}), \eqref{opti-rho} and \eqref{opti-x} represent the offloading choice and channel selection constraints, respectively. Constraints \eqref{opti-k} and \eqref{opti-n} indicate that each UE is assigned to a maximum of one sub-channel and each sub-channel is occupied by a maximum of one UE. $p_{max}$ in \eqref{opti-pmax} is the maximum transmit power constraint, $f_{max}$ in \eqref{opti-fmax} is the maximum GPU frequency of UE$_n$, $t_{max}$ in \eqref{opti-tmax} is the maximum execute latency constraint, $E_n^{max}$ in \eqref{opti-emax} is the battery capacity of UE$_n$, and $\varepsilon_{min}$ in \eqref{opti-amax} is the minimum task accuracy requirement, which is dependent on the task type $\mathcal{M}$ of the current task.

\section{Multi-Agent Proximal Policy Optimization for Resource Allocation}
\label{method}
The binary offloading variable in Constraint \eqref{opti-rho} introduces non-convexity, and the QoE is characterized by a non-convex logistic function, rendering the problem \textbf{P0} a non-convex MINLP problem. To solve the non-convex MINLP problem \textbf{P0}, we propose an advanced semantic-aware MAPPO algorithm, i.e., an actor-critic-based reinforcement learning algorithm, which is suitable for both discrete and continuous action spaces. PPO-based DRL algorithms are known for their stable and reliable updates due to the clipped surrogate objective, which constrains large policy changes during training. This stability is particularly important in communication systems. By contrast, the asynchronous advantage actor-critic (A3C) algorithm uses asynchronous parallel updates without explicit trust-region constraints, which can lead to high variance and unstable learning in multi-agent settings. The multi-agent deep deterministic policy gradient (MADDPG) algorithm, while designed for multi-agent control, relies on centralized action–value critics and deterministic policies that can be sensitive to hyperparameters and suffer from exploration issues in high-dimensional observation spaces.

To utilize the semantic-aware MAPPO algorithm, we reformulate problem \textbf{P0} as an MDP and detail the training process of the MAPPO.

\subsection{Problem Reformulation Based on MDP}
A 4-tuple MDP $\langle\mathcal{S}, \mathcal{A}, \mathcal{R}, \gamma\rangle$ is used to describe the interaction between the ES and the wireless network, where $\mathcal{S}$, $\mathcal{A}$, $\mathcal{R}$ and $\gamma \in \lbrack0,1)$ denote the state space, action space, reward function, and the discount factor, respectively. As agents, UEs need to collect observable states to input into the policy network and obtain their actions. After executing actions, UEs update their observable states and the policy network based on corresponding rewards to improve the output of actions. Specifically, the observable state of UE$_n$ can be represented as $o_n = \left\{g_{k,n}, l_n^U, l_n^S, t_n^U, E_n^U, \varepsilon_n^{\mathcal{M}}, t_{max}, E_n^{max}, \varepsilon_{min}\right\} \in \mathcal{S}$. The action space of UE$_n$ refers to the variables that need to be optimized, which can be expressed as $a_n = \left\{\rho_n, p_n, f_n, \mu_n, x_{k,n}\right\} \in \mathcal{A}$. UEs need to continuously optimize offloading strategies, transmit power, local computation frequency, semantic compression ratio, and channel assignment to purchase better QoE. Drawing from the optimization problem \textbf{P0}, it is evident that the reward function needs to focus on the summation QoE of all UEs. In addition to QoE, the reward function includes normalized penalty terms for latency, energy consumption, and task accuracy, expressed as
\begin{equation}
\begin{aligned}
    r_n = \text{QoE}_n - \left[\frac{t^\mathcal{M}_n-t_{max}}{t_{max}}\right]_{+} - &\left[\frac{E_n^\mathcal{M}-E_n^{max}}{E_n^{max}}\right]_{+} \\- &\left[\frac{\varepsilon_{min}-\varepsilon_n^\mathcal{M}}{\varepsilon_{min}}\right]_{+}, 
\end{aligned}
\end{equation}
where $\left[\cdot\right]_{+} = \max(\cdot,0)$. Therefore, the total reward at the $t$th training timestep is defined as $r_t=\sum^N_{n=1}r_n$. The cumulative long-term reward accounts for the impact of the current state and action on the expected future rewards:
\begin{equation}
    R_t(\tau) = \sum_{t^\prime=t}^T\gamma^{t^\prime-t}r_{t^\prime},
\end{equation}
where $\tau$ is a completed trajectory. Let $\pi_{\boldsymbol{\theta}}$ represents a policy parameterized by $\boldsymbol{\theta}$, where $\boldsymbol{a}_t$ is the collection of $a_n$ at time $t$, and $\boldsymbol{s}_t$ is the collection of $o_n$ at time $t$. Given the state $\boldsymbol{s}_t$, the probability of taking action $\boldsymbol{a}_t$ under policy $\pi_{\boldsymbol{\theta}}$ is denoted as $\operatorname{Pr}(\boldsymbol{a}_t|\boldsymbol{s}_t,\pi_{\boldsymbol{\theta}})$. Consequently, the expected discounted rewards under policy $\pi_{\boldsymbol{\theta}}$, denoted as $J(\pi_{\boldsymbol{\theta}})$, can be expressed as
\begin{equation}
J\left(\pi_{\boldsymbol{\theta}}\right)=\mathbb{E}_{\boldsymbol{a}_t \sim \pi_{\boldsymbol{\theta}}, \boldsymbol{s}_t \sim \mathcal{P}}\left[\sum_{t=1}^{T} \gamma^{t-1} r_t\left(\boldsymbol{s}_t, \boldsymbol{a}_t\right)\right].
\end{equation}
UEs seek to find the optimal policy $\pi_{\boldsymbol{\theta}}$ to maximize the expected cumulative reward $J(\boldsymbol{\pi}_{\boldsymbol{\theta}})$. Therefore, the optimization problem \textbf{P0} is reformulated as an MDP problem:
\begin{equation}
\begin{aligned}
\max _{\pi_{\boldsymbol{\theta}}} \ & J\left(\pi_{\boldsymbol{\theta}}\right) \\
\text { s.t. } \ & \boldsymbol{a}_t \sim \pi\boldsymbol{_\theta}\left(\boldsymbol{a}_t \mid \boldsymbol{s}_t\right), \boldsymbol{s}_{t+1} \sim \operatorname{Pr}\left(\boldsymbol{s}_{t+1} \mid \boldsymbol{s}_t, \boldsymbol{a}_t\right).
\end{aligned}
\end{equation}
\subsection{Semantic-aware MAPPO Algorithm}

\begin{figure}[t]
\centering
\includegraphics[width=0.48\textwidth]{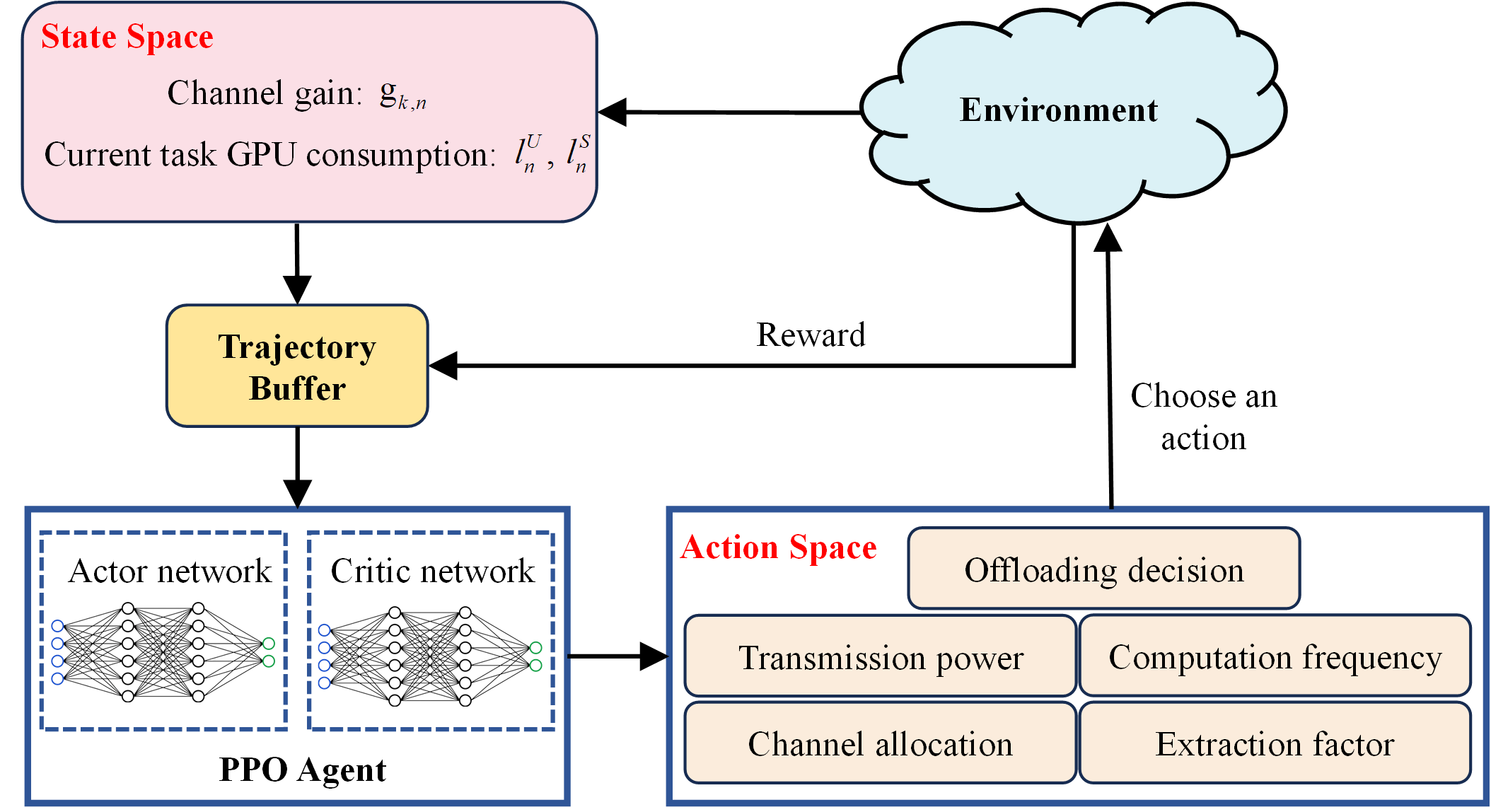}
\captionsetup{justification=raggedright}
\caption{The framework of the MAPPO algorithm.}
\label{RL-model}
\end{figure}

Fig.~\ref{RL-model} illustrates our proposed MAPPO algorithm. PPO is an on-policy algorithm whose trajectories are typically sampled using a fixed policy $\pi_{\boldsymbol{\theta}}$, and policy updates are made using only those trajectories sampled by $\pi_{\boldsymbol{\theta}}$. It utilizes an experience replay buffer, which stores trajectories from past time steps. To enhance environmental exploration efficiency, the MAPPO algorithm employs a parallelized architecture with multiple threads that simultaneously collect environment interaction data and synchronize it during updates. To enhance data efficiency, the MAPPO algorithm typically uses small batches of data for multi-step updates. This leads to a slight discrepancy between the current policy and the policy used during data generation during the update process. The use of importance sampling techniques allows for the assessment of this policy deviation and adjusts the gradient updates, thus stabilizing the learning process. The importance sampling is outlined in \textit{Lemma 1.} \\ \textit{Lemma 1.} Given distributions $x\sim p(x)$ and $x \sim q(x)$, the $\mathbb{E}_{x\sim p}f(x)=\mathbb{E}_{x\sim q}\frac{p(x)}{q(x)}f(x)$.\\ \textit{Proof.} Given $x\sim p(x)$, the expectation of $f(x)$ can be expressed as
\begin{equation}
\begin{aligned}
& \mathbb{E}_{x \sim p} f(x)=\int_x p(x) f(x) \mathrm{d} x \\
& =\int_x \frac{p(x)}{q(x)} q(x) f(x) \mathrm{d} x=\mathbb{E}_{x \sim q} \frac{p(x)}{q(x)} f(x).
\end{aligned}
\end{equation}
Using importance sampling, the optimization function for PPO can be reformulated as
\begin{equation}
J^{\boldsymbol{\theta}^{\prime}}(\boldsymbol{\theta})=\mathbb{E}_{\left(\boldsymbol{s}_t, \boldsymbol{a}_t\right) \sim \pi_{\boldsymbol{\theta}^{\prime}}}\left[\frac{p_{\boldsymbol{\theta}}\left(\boldsymbol{a}_t \mid \boldsymbol{s}_t\right)}{p_{\boldsymbol{\theta}^{\prime}}\left(\boldsymbol{a}_t \mid \boldsymbol{s}_t\right)} A^{\boldsymbol{\theta}^{\prime}}\left(\boldsymbol{s}_t, \boldsymbol{a}_t\right)\right],
\end{equation}
where $\boldsymbol{\theta}$ is the parameter that needs to be optimized while $\boldsymbol{\theta}^\prime$ is the parameter used for sampling. $p_{\boldsymbol{\theta}}\left(\boldsymbol{a}_t \mid \boldsymbol{s}_t\right)$ and $p_{\boldsymbol{\theta}^\prime}\left(\boldsymbol{a}_t \mid \boldsymbol{s}_t\right)$ represent taking action $\boldsymbol{a}_t$ given state $\boldsymbol{s}_t$ with $\boldsymbol{\theta}$ and $\boldsymbol{\theta}^\prime$, respectively. $\boldsymbol{\theta}^\prime$ interacts with the environment to generate trajectories while $\boldsymbol{\theta}$ does not interact with the environment directly but learns from the trajectories sampled by $\boldsymbol{\theta}^\prime$. The advantage function $A^{\boldsymbol{\theta}^{\prime}}\left(\boldsymbol{s}_t, \boldsymbol{a}_t\right)$  represents the advantage of state-action pairs $\left(\boldsymbol{s}_t, \boldsymbol{a}_t\right)$ obtained through the interaction of $\boldsymbol{\theta}^{\prime}$ with the environment, which is defined as
\begin{equation}
\label{state-action-value}
A^{\boldsymbol{\theta}^\prime}\left(\boldsymbol{s}_t, \boldsymbol{a}_t\right)=Q^{\boldsymbol{\theta}^\prime}\left(\boldsymbol{s}_t, \boldsymbol{a}_t\right)-V^{\boldsymbol{\phi}}\left(\boldsymbol{s}_t\right),
\end{equation}
where $\boldsymbol{\phi}$ is the parameters of critic network.
The state-action value function $Q^{\boldsymbol{\theta}^\prime}\left(\boldsymbol{s}_t, \boldsymbol{a}_t\right)$ is represented as
\begin{equation}
    Q^{\boldsymbol{\theta}^\prime}\left(\boldsymbol{s}_t, \boldsymbol{a}_t\right) = \mathbb{E}_{\boldsymbol{a}_t \sim \pi_{\boldsymbol{\theta^{\prime}}},\boldsymbol{s}_t \sim \mathcal{P}}\left[\sum_{t^{\prime}=t}^T\gamma^{t^{\prime}-t}r_{t^{\prime}} \right],
\end{equation}
and the state-value function is defined as
\begin{equation}
    V^{\boldsymbol{\phi}}\left(\boldsymbol{s}_t \right) = \mathbb{E}_{\boldsymbol{s}_t \sim \mathcal{P}}\left[\sum_{t^{\prime}=t}^T\gamma^{t^{\prime}-t}r_{t^{\prime}} \right].
\end{equation}
It is noted that $Q^{\boldsymbol{\theta}^\prime}\left(\boldsymbol{s}_t, \boldsymbol{a}_t\right)$ can also be expressed by temporal difference (TD) form as $Q^{\boldsymbol{\theta}^\prime}\left(\boldsymbol{s}_t, \boldsymbol{a}_t\right) = r_t + \gamma V^{\boldsymbol{\phi}}\left(\boldsymbol{s}_{t+1} \right)$.
Therefore, \eqref{state-action-value} can be reformulated as
\begin{equation}
    A^{\boldsymbol{\theta}^\prime}\left(\boldsymbol{s}_t, \boldsymbol{a}_t\right)=r_t + \gamma V^{\boldsymbol{\phi}}\left(\boldsymbol{s}_{t+1} \right)-V^{\boldsymbol{\phi}}\left(\boldsymbol{s}_t \right).
\end{equation}
However, estimating based on TD often carries a higher bias. Thus, we employ Generalized Advantage Estimation (GAE) \cite{schulman2015high} to evaluate the current state-action value function. The GAE advantage function can be expressed as
\begin{equation}
\label{GAE}
    A^{\mathrm{GAE}}(\boldsymbol{s}_t, \boldsymbol{a}_t)=\sum_{l=0}^{T-t}(\lambda\gamma)^l\delta_{t+l},
\end{equation}
where $\delta_{t}=A^{\boldsymbol{\theta}^\prime}\left(\boldsymbol{s}_t, \boldsymbol{a}_t\right)$. $\lambda \in (0,1]$ is a discount factor. Note that \eqref{GAE} aims to balance the variance and bias of the advantage estimates by leveraging information from multiple future steps while discounting their influence.

If there are no constraints on the policies $\pi_{\boldsymbol{\theta}}$ and $\pi_{\boldsymbol{\theta}^\prime}$, significant fluctuations may occur during policy updates. This is due to the limited sampling, which fails to accurately represent the probability distributions described in \textit{Lemma 1}. To constrain the divergence between policies $\pi_{\boldsymbol{\theta}}$ and $\pi_{\boldsymbol{\theta}^\prime}$, $J^{\mathrm{CLIP}}(\boldsymbol{\theta})$ is employed. It is defined as
\begin{equation}
\begin{aligned}
    J^{\mathrm{CLIP}}(\boldsymbol{\theta})
= \mathbb{E}_t \Big[
    \min \Big(
        r_t(\boldsymbol{\theta})
        A^{\mathrm{GAE}}
        (\boldsymbol{s}_t, \boldsymbol{a}_t), \\
        \qquad
        g\!\left(
            \epsilon,
            A^{\mathrm{GAE}}
            (\boldsymbol{s}_t, \boldsymbol{a}_t)
        \right)
    \Big)
\Big],
\end{aligned}
\end{equation}
where $r_t(\boldsymbol{\theta})=\frac{p_{\boldsymbol{\theta}}\left(\boldsymbol{a}_t \mid \boldsymbol{s}_t\right)}{p_{\boldsymbol{\theta}^{\prime}}\left(\boldsymbol{a}_t \mid \boldsymbol{s}_t\right)}$ and $g$ is the clip function which can be defined as
\begin{equation}
\begin{aligned}
    g(\epsilon, &A^{\boldsymbol{\theta}^\prime}\left(\boldsymbol{s}_t, \boldsymbol{a}_t\right))=\\
    &\begin{cases}(1+\epsilon) A^{\mathrm{GAE}}\left(\boldsymbol{s}_t, \boldsymbol{a}_t\right),\ \text{if} \ A^{\mathrm{GAE}}\left(\boldsymbol{s}_t, \boldsymbol{a}_t\right) \geq 0, \\ (1-\epsilon) A^{\mathrm{GAE}}\left(\boldsymbol{s}_t, \boldsymbol{a}_t\right),\ \text{if} \ A^{\mathrm{GAE}}\left(\boldsymbol{s}_t, \boldsymbol{a}_t\right)<0,\end{cases}
\end{aligned}
\end{equation}
where $\epsilon$ is a hyperparameter, which can control the step size of the policy network updates.

\begin{figure}[t]
\centering
\subfloat[$A\geq0$.\label{j-clip-A+}]{\includegraphics[width=0.23\textwidth]{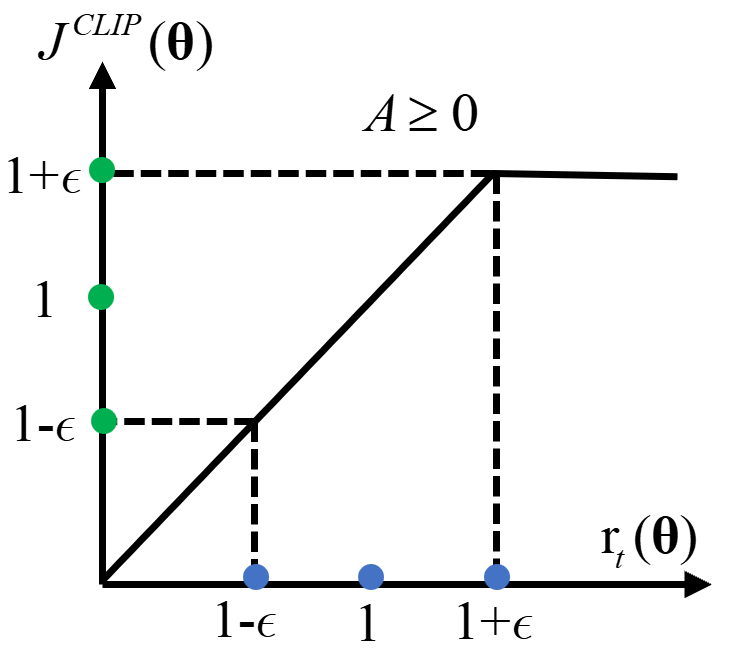}}
\hfil
\subfloat[$A<0$.\label{j-clip-A-}]{\includegraphics[width=0.23\textwidth]{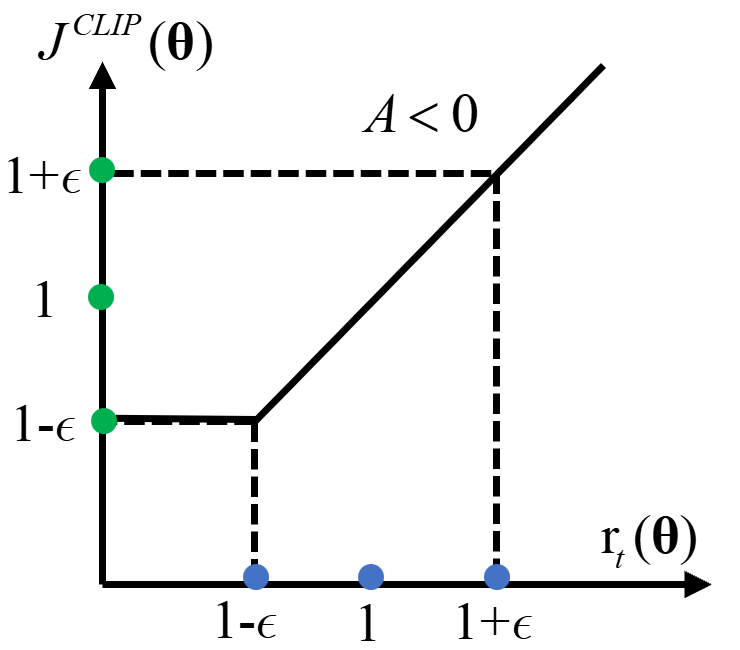}}
\captionsetup{justification=raggedright}
\caption{The impact of $A$ on $J^{CLIP}$.}
\label{j-clip}
\end{figure}
Fig.~\ref{j-clip} illustrates the impact of $A$ on the $J^{\mathrm{CLIP}}$ function. Fig.~\ref{j-clip-A+} and Fig.~\ref{j-clip-A-} correspond to cases where $A\geq0$ and $A<0$, respectively. When $A\geq0$, the state-action pair is favorable, and we aim to increase its probability; that is, the larger $p_{\boldsymbol{\theta}}$ is, the better. However, the maximum value of $r_t(\boldsymbol{\theta})$ cannot exceed $1+\epsilon$, meaning the ratio of $p_{\boldsymbol{\theta}}$ to $p_{\boldsymbol{\theta^{\prime}}}$ should not exceed $1+\epsilon$. When $A<0$, the state-action pair is unfavorable, and we aim to reduce $p_{\boldsymbol{\theta}}$, but the ratio of $p_{\boldsymbol{\theta}}$ to $p_{\boldsymbol{\theta^{\prime}}}$ should not fall below $1-\epsilon$. By using the CLIP function, the difference between $p_{\boldsymbol{\theta}}$ and $p_{\boldsymbol{\theta^{\prime}}}$ is moderate, which helps improve the model's convergence speed and performance.
The loss function of the critic network aims to minimize the discrepancy between the state-value function and the cumulative reward and is expressed as
\begin{equation}
L_t^{\mathrm{CL}}=\left\|{V}^{\boldsymbol{\phi}}\left(\boldsymbol{s}_{t}\right)-R_t(\tau)\right\|^2.
\end{equation}
To enhance the stochasticity of the policy, the ultimate optimization object for the MAPPO algorithm is formulated as
\begin{equation}
\begin{aligned}
    L_t^{\mathrm{CL+CLIP+S}}(\theta)
= \mathbb{E}_t \Big[
    & J_t^{\mathrm{CLIP}}
     - b_1 L_t^{\mathrm{CL}}\\
    & + b_2 H\!\left(
        \pi_\theta(\cdot \mid s_t)
      \right)
\Big],
\end{aligned}
\end{equation}
where $b_1$ and $b_2$ are weight coefficients. $H$ is the entropy bonus to ensure sufficient exploration.

Actor-critic based PPO algorithm comprises a policy network $\boldsymbol{\theta}$ for outputting actions and a critic network $\boldsymbol{\phi}$. The parameters $\boldsymbol{\theta}$ are updated by
\begin{equation}
\label{update_theta}    \text{arg}\max_{\boldsymbol{\theta}}\hat{\mathbb{E}}_t\left[J_t^{\mathrm{CLIP}}+b_2 H(\pi_{\boldsymbol{\theta}}\left(\cdot\mid \boldsymbol{s}_t\right))\right],
\end{equation}
and parameters $\boldsymbol{\phi}$ are updated by
\begin{equation}
\label{update_phi}    \text{arg}\min_{\boldsymbol{\phi}}\hat{\mathbb{E}}_t\left[L_t^{\mathrm{CL}}\right].
\end{equation}
The expectation $\hat{\mathbb{E}}_t$ is estimated using data obtained from interactions between the agent and the environment. Monte Carlo sampling is utilized to approximate the expectation $\hat{\mathbb{E}}$.

\begin{algorithm}[t] \footnotesize
	\caption{MAPPO Training Algorithm}
    \textbf{Initialization:} Initialize the policy network with parameters $\boldsymbol{\theta}$, the sampling policy network with parameters $\boldsymbol{\theta^\prime}$ and the critic network with parameters $\boldsymbol{\phi}$. Initialize the observation space $o_n$ and the action space $a_n$.\\
	\textbf{Input:} Corresponding channel gain $g_{k,n}$ and task queue $Q_n$.\\
	\textbf{Output:} Trained neural network.
    \begin{algorithmic}[1]
    \FOR{\textit{episodes}}
        \STATE $\pi_{\boldsymbol{\theta^\prime}} \leftarrow \pi_{\boldsymbol{\theta}}$;
        \STATE Run policy $\pi_{\boldsymbol{\theta^\prime}}$ for $T$ steps and save the trajectory $\tau$ to the replay buffer;
        \STATE Calculate $T$ step GAE $A^{\mathrm{GAE}}_1,...,A^{\mathrm{GAE}}_T$ using $\pi_{\boldsymbol{\theta^\prime}}$ and $\boldsymbol{\phi}$;
        \FOR{\textit{epochs}}
            \STATE Update $\boldsymbol{\theta}$ using \eqref{update_theta};
            \STATE Update $\boldsymbol{\phi}$ using \eqref{update_phi};
        \ENDFOR
    \ENDFOR
    \end{algorithmic}
\end{algorithm}

\section{Simulation Results}
\label{simulation}
In this section, the performance of the semantic-aware multi-task offloading model will be demonstrated. We provide specific implementation details, investigate the QoE performance of the proposed approach, and examine the impact of user preferences on the system.
\subsection{Implementation Details}
Three tasks are considered: text classification, image classification, and VQA, corresponding to text, image, and multi-modal tasks, respectively. For text classification, we use the SST-2 dataset, for image classification, the CIFAR-10 dataset, and for VQA, the VQAv2 dataset is adopted. The text semantic extraction network utilizes an advanced BERT-based \cite{devlin2018bert} text embedding model, while the image semantic extraction network is based on the Vision Transformer (ViT) \cite{vit2021}. For VQA, the encoder architecture employs BERT for linguistic processing and ViT for visual feature extraction, with cross-modal fusion and answer generation performed at the receiver. The text classification model requires 3.72 GFLOPs, the image classification model requires 8.43 GFLOPs, and the VQA model requires 8.31 GFLOPs. Training is conducted using the AdamW optimizer, with a learning rate of $3\times10^{-5}$, a batch size of 50, and a weight decay of $1\times10^{-4}$. All experiments were conducted on a platform equipped with an NVIDIA RTX 4090 and a CPU Intel i9-13900K @5.8 GHz. When the training of all semantic transmission models is complete, we can calculate the GPU consumption for basic semantic extraction $l_n^{SE}$ and the GPU consumption $l^S$ for execution at the ES. 

The architecture of MAPPO is detailed in Table \ref{tab:mappo_arch}. The actor network takes the state information as input and first extracts intermediate representations through a two-layer multilayer perceptron (MLP) with 256 hidden units per layer. The resulting features are then fed into a multi-layer gated recurrent unit (GRU) network to capture temporal dependencies and generate action-related features. Finally, an additional MLP layer maps these features to the action probability distribution. Similarly, the critic network takes the state information as input and processes it through a two-layer MLP with 256 hidden units to obtain intermediate features. These features are subsequently passed to a multi-layer GRU network to produce value-related representations, which are then mapped to the state value via a final MLP layer.

Each UE has a queue length of 20, containing tasks from different task types. Additional details regarding the wireless environment, GPU computational frequency, and the proposed semantic-aware MAPPO algorithm training configurations are summarized in Table \ref{parameters}. 

In our experiments, the model sizes for different modality-specific tasks range from several tens to a few hundred megabytes, while the computational complexity of the models is approximately 3 to 9 GFLOPs. Taking a commonly used edge computing platform, the NVIDIA Jetson Nano, as an example, which provides a peak computing capability of about 470 GFLOPS (Floating Point Operations per Second), the proposed models can be efficiently deployed and executed under the given computational constraints. Even for devices with more limited computing resources, the proposed framework is fully compatible with standard model optimization techniques, including model pruning, weight quantization (e.g., 8-bit or mixed-precision inference), and knowledge distillation. Furthermore, selecting an appropriate semantic compression factor $\mu$ can also be tailored to the computational capabilities of the current device.

To verify the effectiveness of the semantic-aware multi-task offloading framework, we consider the following benchmarks:

\begin{table}[t]
\centering

\caption{Network Architectures of MAPPO Actor and Critic}
\begin{tabular}{|m{1.5cm}<{\centering} |m{3cm}<{\centering} |m{2.5cm}<{\centering}|}
\hline
\textbf{Network Component} & \textbf{Layer Description} & \textbf{Output / Notes} \\ \hline
\multirow{4}{*}{Actor} 
& Fully-Connected (256 units), ReLU, LayerNorm & Intermediate feature \\ \cline{2-3}
& Fully-Connected (256 units), ReLU, LayerNorm & Refined feature \\ \cline{2-3}
& Multi-layer GRU & Action feature \\ \cline{2-3}
& Fully-Connected & Action probability distribution \\ \hline
\multirow{4}{*}{Critic}
& Fully-Connected (256 units), ReLU, LayerNorm & Intermediate feature \\ \cline{2-3}
& Fully-Connected (256 units), ReLU, LayerNorm & Refined feature \\ \cline{2-3}
& Multi-layer GRU & Value feature \\ \cline{2-3}
& Fully-Connected & State value estimate \\ \hline
\end{tabular}
\label{tab:mappo_arch}
\end{table}

\begin{table}[t]
\centering
\caption{Simulation Parameters}
\begin{tabular}{|l|c|}
\hline
\textbf{Parameters} & \textbf{Values} \\
\hline
Number of UEs $N$ & 4 \\\hline
Sub-channel bandwidth $B$ & 10 MHz \\\hline
Task queue length $Q_n$ & 20 \\ \hline
Noise power $\sigma_z^2$ & 2 mW \\\hline
Transmit power $p_n$ range & (10, 90) mW \\\hline
Semantic extraction factor $\mu_n$ & (0.1, 1) \\ \hline
The number of user CUDA cores & 1280 \\ \hline
Local GPU frequency $f_n$ range & (1.5, 1.7) GHz \\\hline
The number of ES CUDA cores & 65536 \\ \hline
Remote GPU frequency $f_s$ & 2.2 GHz \\\hline
Energy consumption coefficient of UEs $\alpha^U$ & $1\times10^{-26}$ \\\hline
Energy consumption of ES $\alpha^S$ & $1.2\times10^{-26}$ \\\hline
Minimum accuracy requirements $\epsilon_{\min}$ & 50\% \\\hline
Execution latency constraint $t_{\max}$ & 5 ms \\\hline
Training episode of the MAPPO algorithm & 300 \\\hline
Testing episode of the MAPPO algorithm & 200 \\\hline
Batch size of training the MAPPO & 2 \\\hline
Epoch of training the MAPPO & 5 \\\hline
Learning rate & $1\times10^{-4}$ \\\hline
Advantage discount factor $\lambda$ & 0.95 \\\hline
Reward discount factor $\gamma$ & 0.99 \\\hline
PPO-Clip parameter $\epsilon$ & 0.2 \\\hline
Policy entropy bonus weight $b_2$ & 0.1 \\\hline
The slope of logistic function $\lambda, \beta, \eta$ & $1\times10^3$, 10, 2\\
\hline
\end{tabular}
\label{parameters}
\end{table}

\begin{itemize}
\item \textbf{Semantic-aware MAPPO (discrete):} This benchmark discretizes continuous actions in the hybrid action space, while keeping other settings the same as MAPPO.
\item \textbf{Semantic-unaware MAPPO:} The benchmark without semantic extraction during offloading, i.e. $\mu_n=1$. Specifically, this scheme relies on conventional compression techniques and does not support adaptive semantic extraction. All other settings are identical to those used in semantic-aware MAPPO.
\item \textbf{Semantic-aware Dueling Double DQN (D3QN):} D3QN can only handle discrete action spaces, so we also discretized the action space and used the same environment as semantic-aware MAPPO.
\item \textbf{Semantic-unaware D3QN:} The benchmark without semantic extraction during offloading, i.e. $\mu_n=1$. All other settings are identical to those used in semantic-aware D3QN.
\item \textbf{Local:} All tasks are executed locally.
\end{itemize}
All experimental results are averaged over 200 test runs to mitigate the effects of uneven task type distribution and dynamic changes in the channel.
\subsection{QoE Performance and Convergence}
\begin{figure}[t]
\centering
\includegraphics[width=0.48\textwidth]{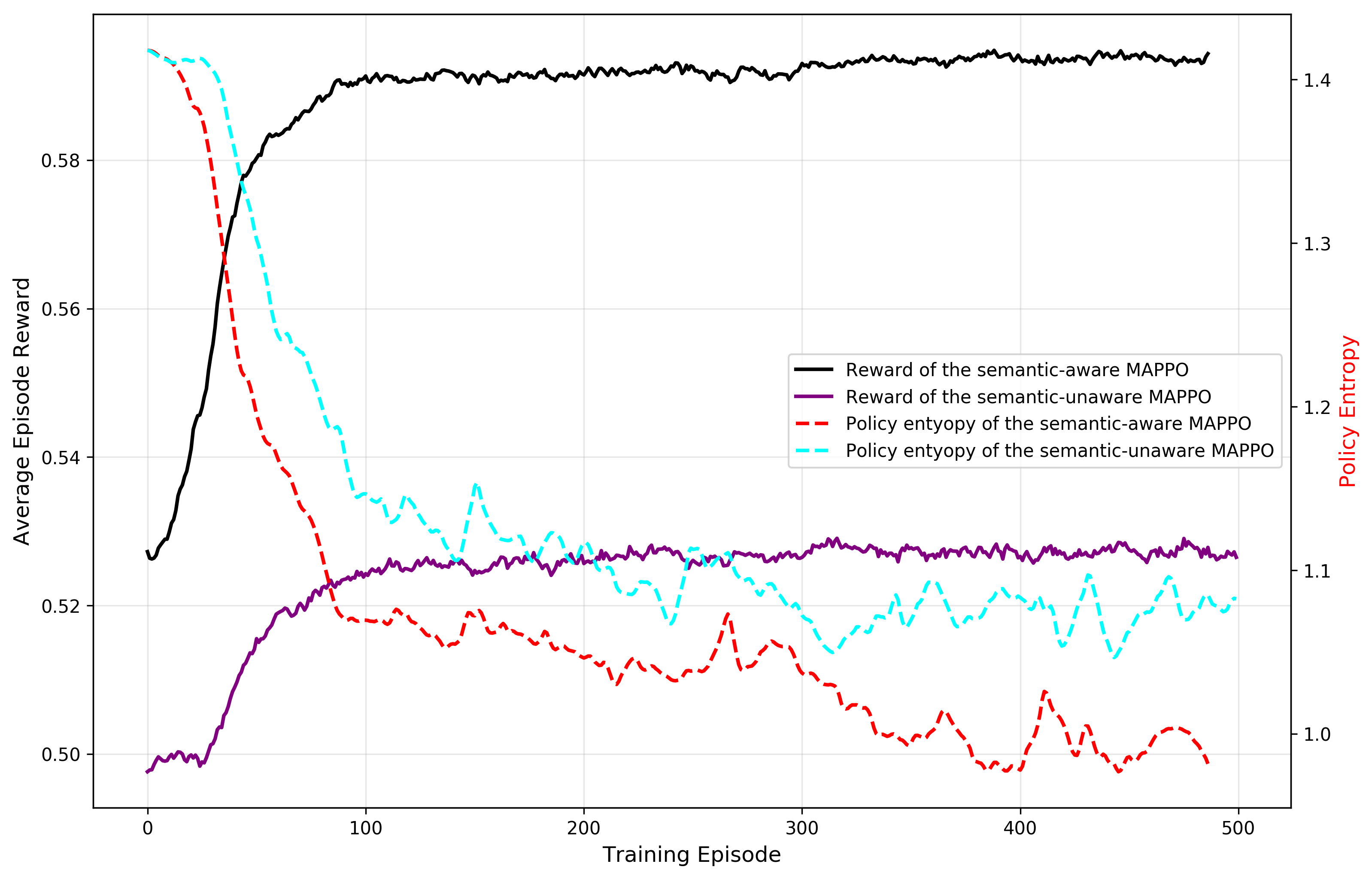}
\captionsetup{justification=raggedright}
\caption{Convergence of the MAPPO reward.}
\label{MAPPO_convergence}
\end{figure}

\textit{1) Convergence Performance:} The preference parameters, $\omega_t$, $\omega_e$, and $\omega_a$ are all set to 1/3, indicating that the model has no specific preference among latency, energy consumption, and accuracy. Fig.~\ref{MAPPO_convergence} shows that the proposed MAPPO algorithm converges after approximately 100 episodes and maintains optimal performance. The policy entropy also gradually converges, indicating that the policy becomes stable during training. For the D3QN algorithm, the replay buffer size is set to 30000. 

\begin{figure}[t]
\centering
\includegraphics[width=0.48\textwidth]{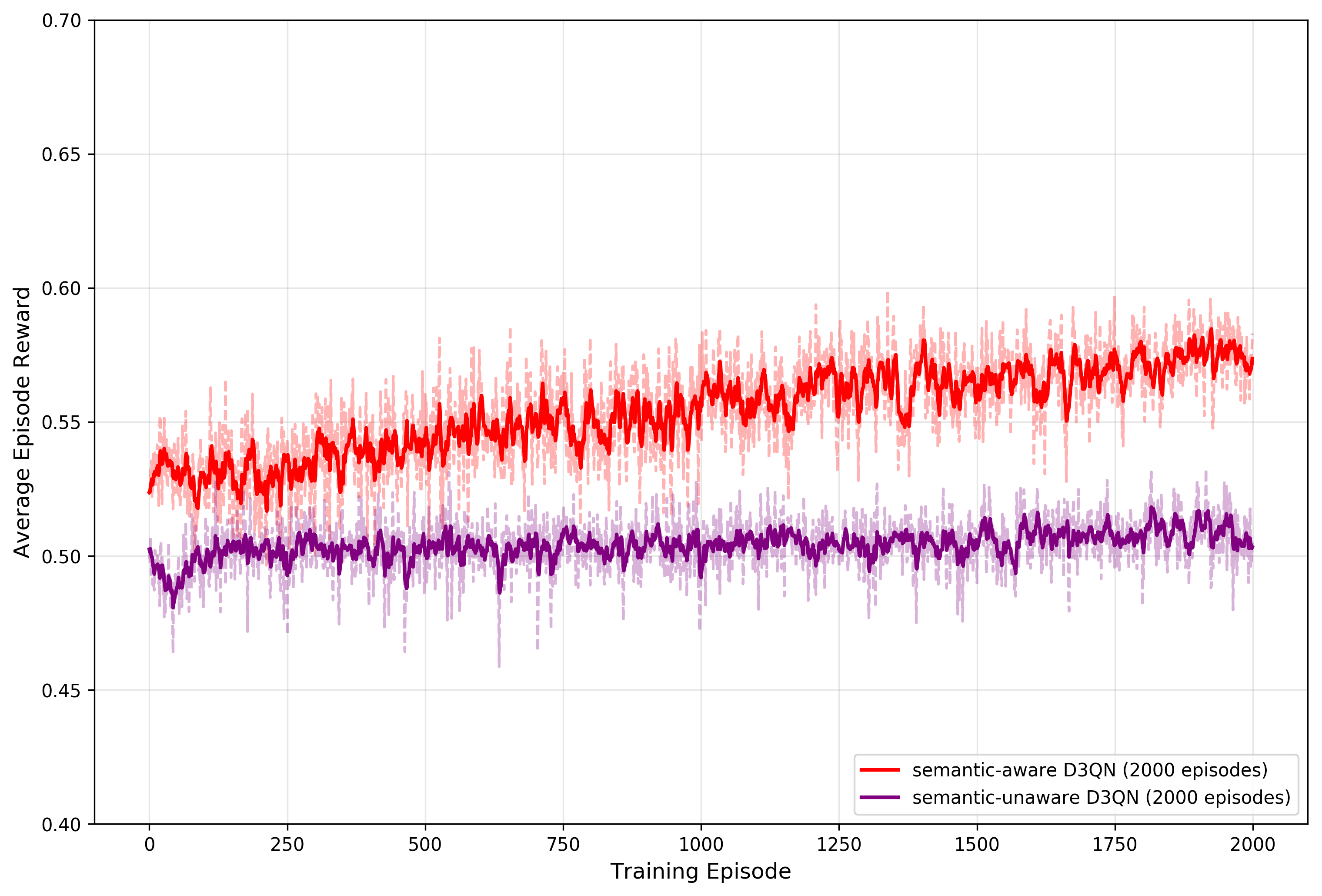}
\captionsetup{justification=raggedright}
\caption{Convergence of the D3QN reward.}
\label{D3QN_convergence}
\end{figure}

Fig.~\ref{D3QN_convergence} shows that the D3QN algorithm finally converges after 2000 episodes. The proposed MAPPO algorithm directly optimizes cumulative returns through policy gradient methods, while employing a trust region mechanism to constrain policy update magnitudes, thereby preventing training instability caused by abrupt policy changes. In contrast, the D3QN algorithm relies on TD learning and demonstrates lower learning efficiency in environments with sparse rewards or delayed feedback. Furthermore, to enhance the sampling efficiency of on-policy algorithms, the proposed MAPPO algorithm employs multi-threaded asynchronous sampling and completes state collection and synchronization during network updates. Consequently, the proposed MAPPO algorithm achieves faster convergence with greater stability compared to D3QN.




\begin{figure}[t]
\centering
\includegraphics[width=0.48\textwidth]{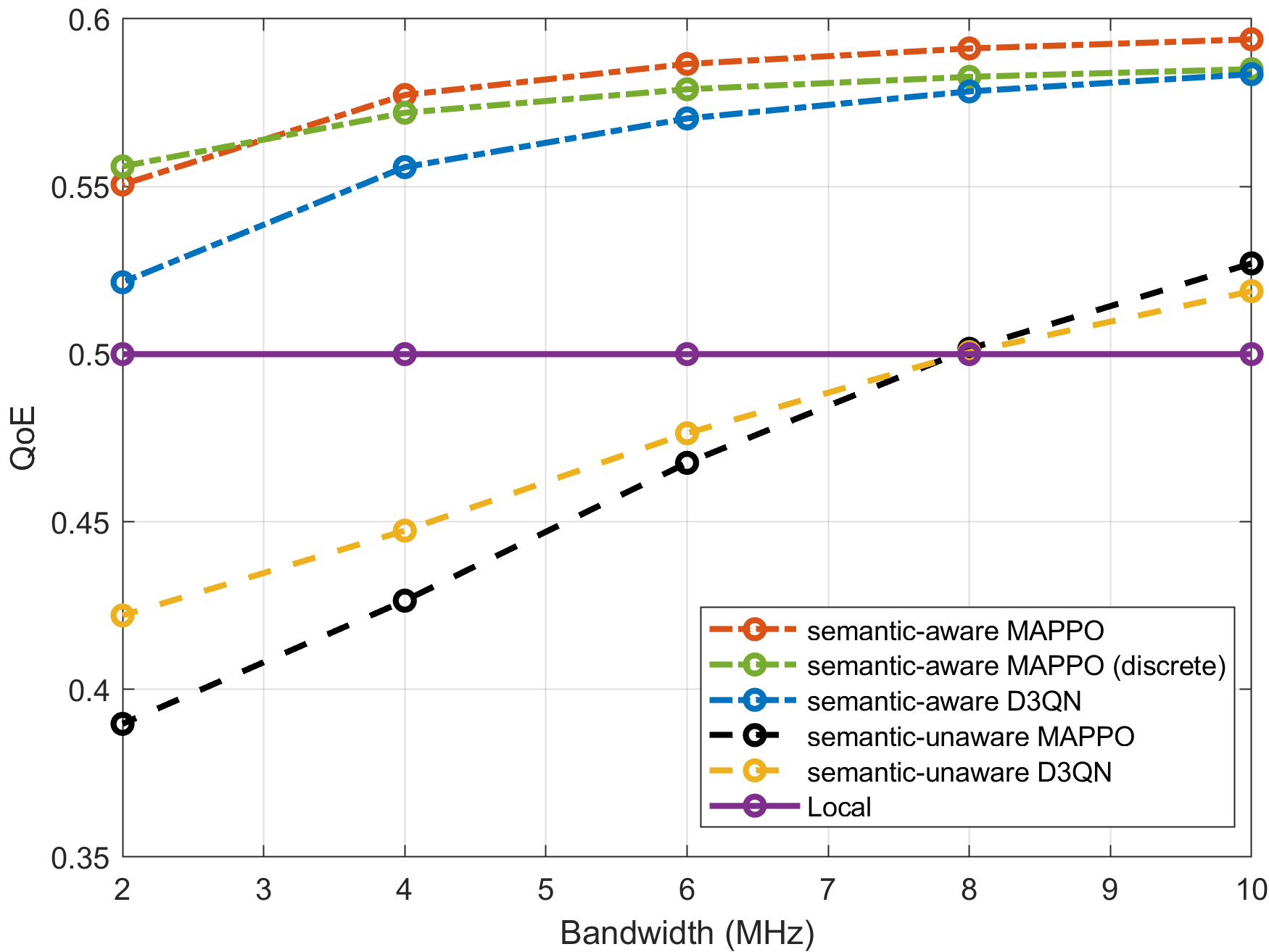}
\captionsetup{justification=raggedright}
\caption{QoE with different sub-channel bandwidths.}
\label{qoe_bandwidth}
\end{figure}

\textit{2) QoE Performance:} Fig.~\ref{qoe_bandwidth} reveals the relationship between QoE and the sub-channel bandwidth for UEs. All schemes are initially trained at a sub-channel bandwidth of 10 MHz and then tested under various bandwidth conditions. It is observed that the proposed semantic-aware MAPPO algorithm achieves the best performance across all bandwidth conditions. Comparative evaluations demonstrate 12.68\% QoE improvements versus semantic-unaware MAPPO and 14.48\% superiority versus semantic-unaware D3QN. Semantic-unaware methods fix $\mu_n=1$ without semantic compression, increasing transmission and execution latency, consequently degrading QoE. The D3QN algorithm requires discrete action space approximation, which may introduce quantization errors or reduce sampling efficiency. The semantic-aware MAPPO (discrete) method adopts the same discrete action space configuration as D3QN. In contrast, the proposed MAPPO algorithm directly outputs continuous action probability distributions, and support hybrid action space, eliminating the need for discretization and thereby avoiding the curse of dimensionality in high-dimensional action spaces. Consequently, our method achieves superior QoE. Local execution is solely related to the local computation frequency. When calculating QoE, the local computation frequency is fixed, meaning we use the local execution’s latency, energy consumption, and accuracy to normalize the latency, energy, and accuracy obtained from other methods. Thus, the QoE of local execution is fixed at 0.5.

\begin{figure}[t]
\centering
\includegraphics[width=0.48\textwidth]{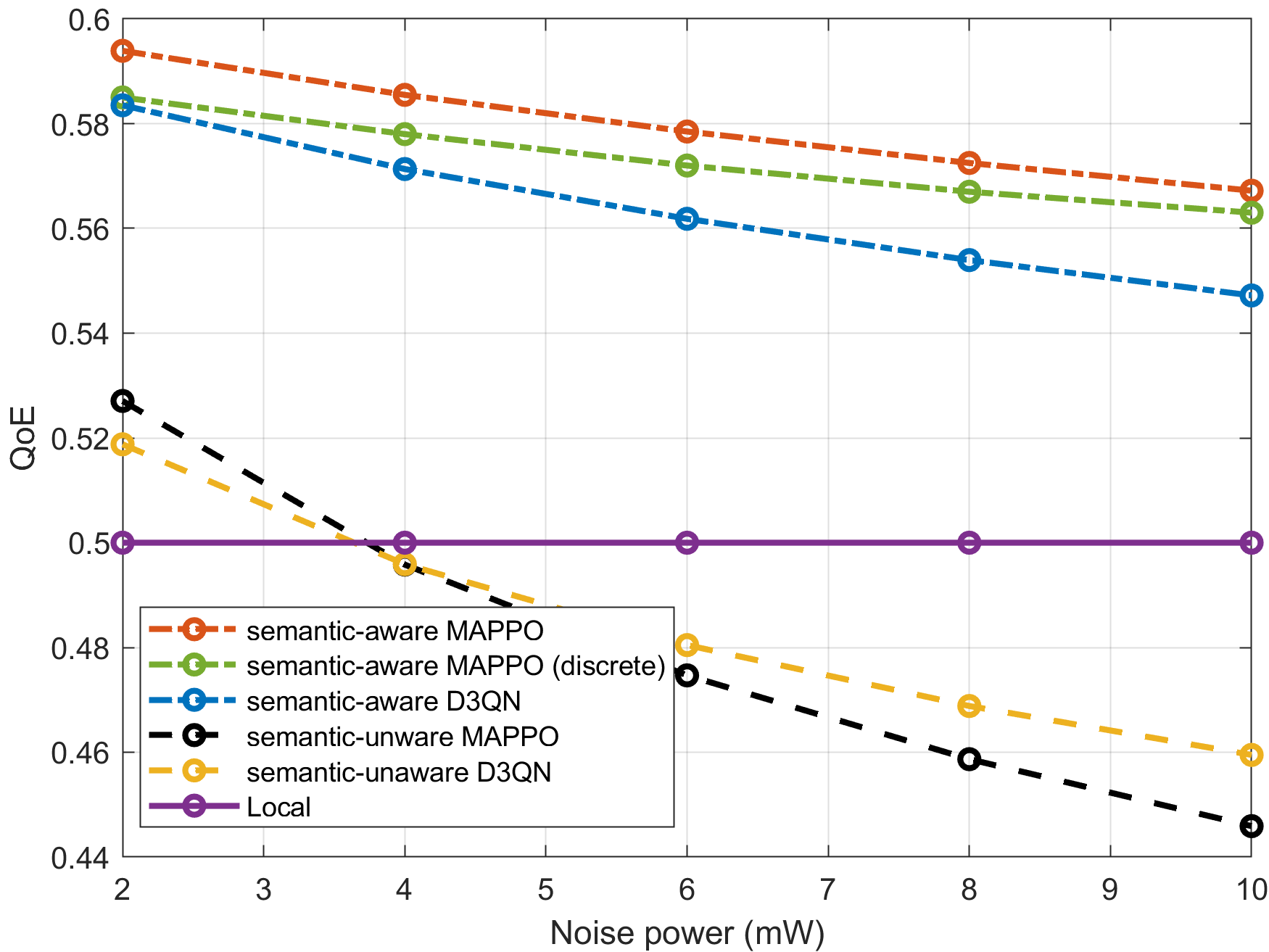}
\captionsetup{justification=raggedright}
\caption{QoE with different noise powers.}
\label{qoe_noise}
\end{figure}

\begin{figure}[t]
\centering
\includegraphics[width=0.48\textwidth]{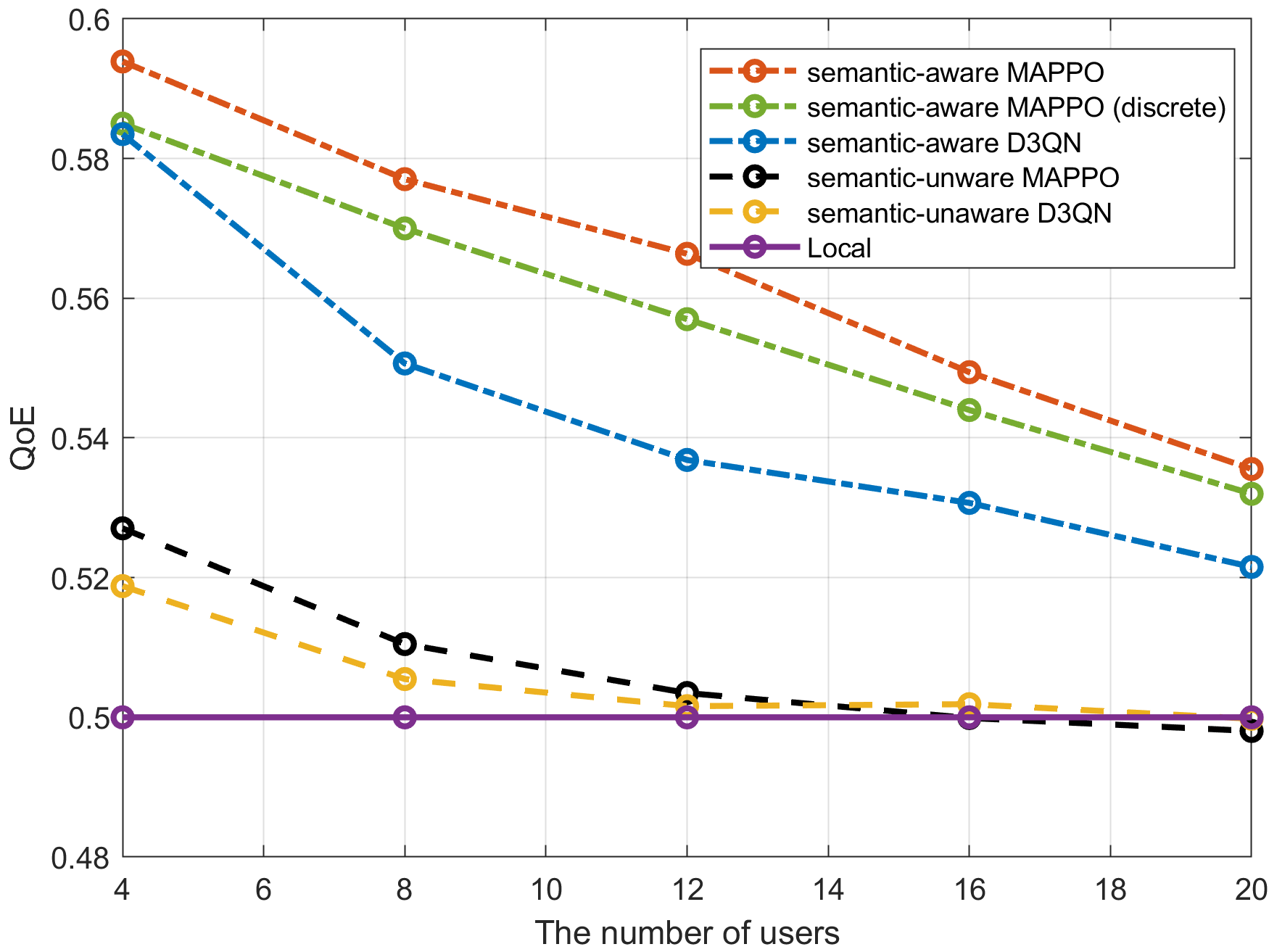}
\captionsetup{justification=raggedright}
\caption{QoE with different numbers of users.}
\label{qoe_num_agent}
\end{figure}

\begin{figure}[t]
\centering
\includegraphics[width=0.48\textwidth]{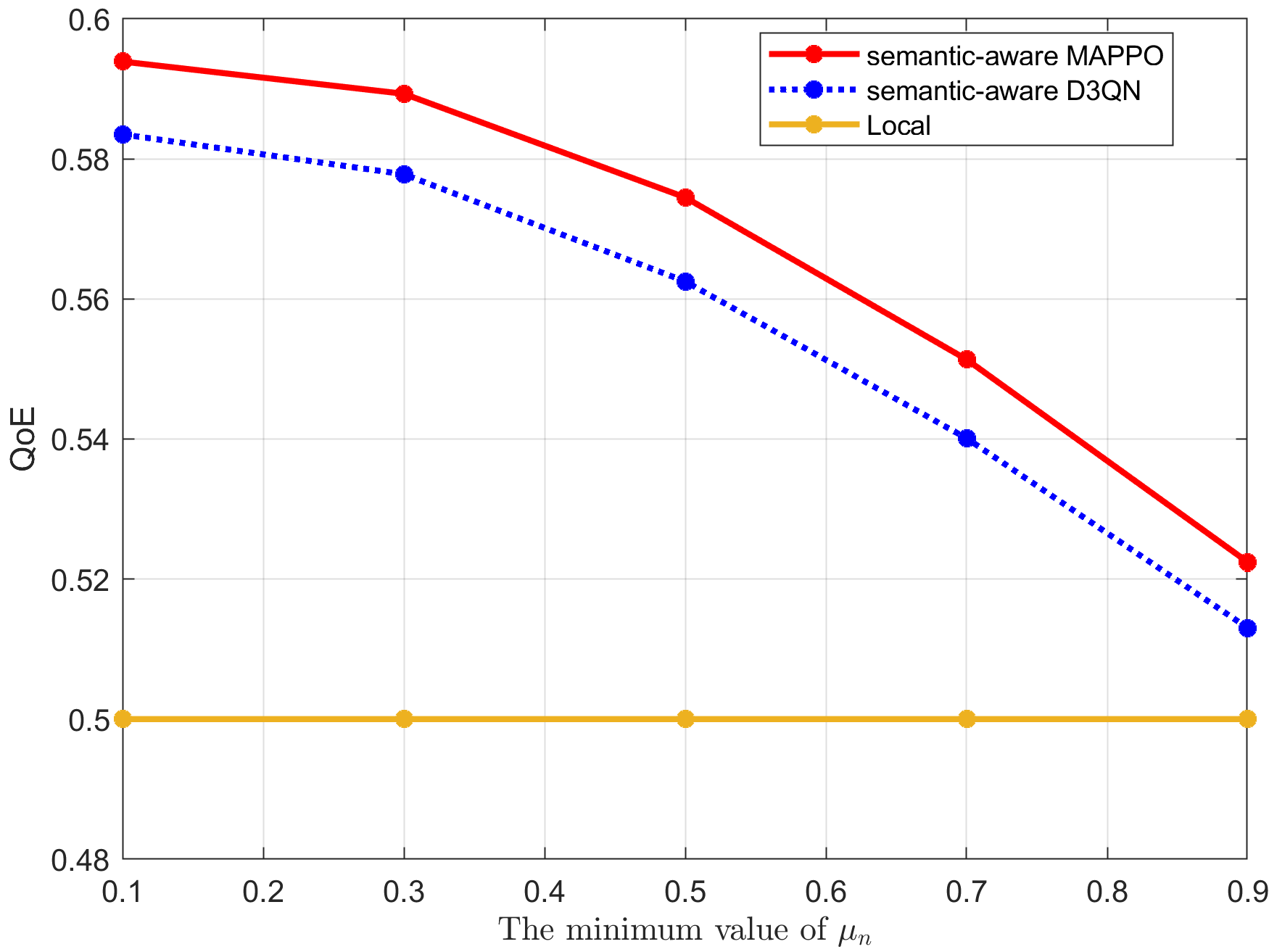}
\captionsetup{justification=raggedright}
\caption{QoE with different minimum semantic extraction factor $\mu_n$.}
\label{qoe_mu}
\end{figure}
Fig.~\ref{qoe_noise} describes how QoE varies with channel noise intensity. All schemes are first trained under the noise intensity of 2 mW and then tested at different noise intensities. Similarly, the proposed semantic-aware MAPPO algorithm performs better than all other benchmarks. The semantic-aware algorithm dynamically selects optimal semantic extraction factors based on real-time noise intensity and channel gain, maintaining stable QoE without significant degradation even under increasing noise levels. In contrast, semantic-unaware schemes lack adaptive semantic symbol quantity regulation, resulting in substantial performance deterioration. Fig.~\ref{qoe_num_agent} illustrates the QoE trend versus the number of users. As the number of users increases, the server must handle more computation offloading requests, leading to elevated processing latency and energy consumption, which collectively degrade overall QoE. Semantic-unaware approaches cannot leverage semantic extraction to reduce uplink latency during user scaling, ultimately forcing all users to execute tasks locally. Consequently, its QoE converges to the local computing baseline. Fig.~\ref{qoe_mu} illustrates the QoE variation versus different minimum $\mu_n$. We retrain the MAPPO under different minimum values of $\mu_n$. Results demonstrate that when users have fewer selectable semantic extraction factors, the inability to utilize smaller compression factors for data size reduction significantly impairs channel adaptation under poor conditions, ultimately degrading QoE.
\begin{table}[t]
\centering
\caption{Performance under Different Preferences}
\begin{tabular}{|c|c|c|c|}
\hline
& Delay (ms) & \makecell{Energy \\ Consumption (J)} & Accuracy (\%)\\ \hline
No Preference & 0.523 & 0.0186 & 51.37\\ \hline
Delay Preference & \textbf{0.369} & 0.0192 & 43.08 \\ \hline
Energy Preference & 3.50 & \textbf{0.0163} & 56.85 \\ \hline
Accuracy Preference & 1.67 & 0.0682 & \textbf{71.99} \\ \hline
\end{tabular}
\label{different_preference}
\end{table}

\begin{figure}[t]
\centering
\includegraphics[width=0.48\textwidth]{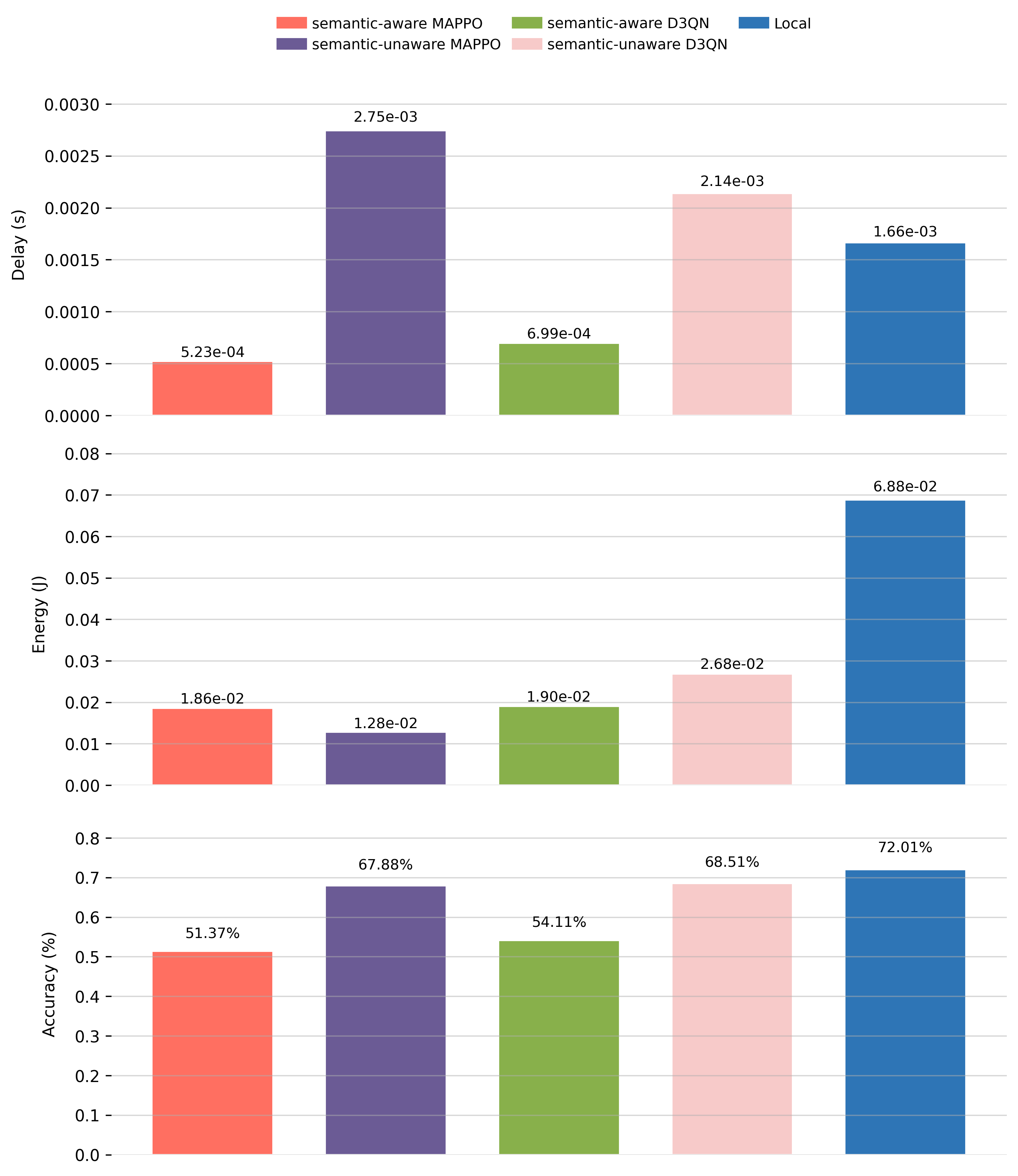}
\captionsetup{justification=raggedright}
\caption{Performance of different metrics.}
\label{performance_metrics}
\end{figure}

To further investigate the contributions of latency, energy consumption, and task performance to QoE, Fig.~\ref{performance_metrics} illustrates the execution latency, energy consumption, and task accuracy. All metrics are measured under the following conditions: 10 MHz bandwidth, 2 mW noise power, and 4 users. Local task execution demonstrates longer latency and higher energy consumption, yet achieves optimal task performance. This occurs because local processing avoids performance degradation induced by semantic compression. However, the prolonged execution time incurs significant energy overhead. Semantic-unaware approaches exhibit inferior latency performance because of the transmission of raw task data without semantic extraction, resulting in extended transmission delays. However, server-side processing of offloaded tasks requires minimal time, yielding lower energy consumption compared to local execution. Owing to wireless channel effects during task offloading, semantic-unaware methods show slightly reduced task accuracy relative to local processing. The semantic-aware approach dynamically selects optimal semantic extraction factors based on real-time channel conditions, computational load, and transmission requirements, achieving the best trade-off among latency, energy efficiency, and task accuracy. Consequently, it delivers the highest QoE.

\begin{figure}[t]
\centering
\includegraphics[width=0.48\textwidth]{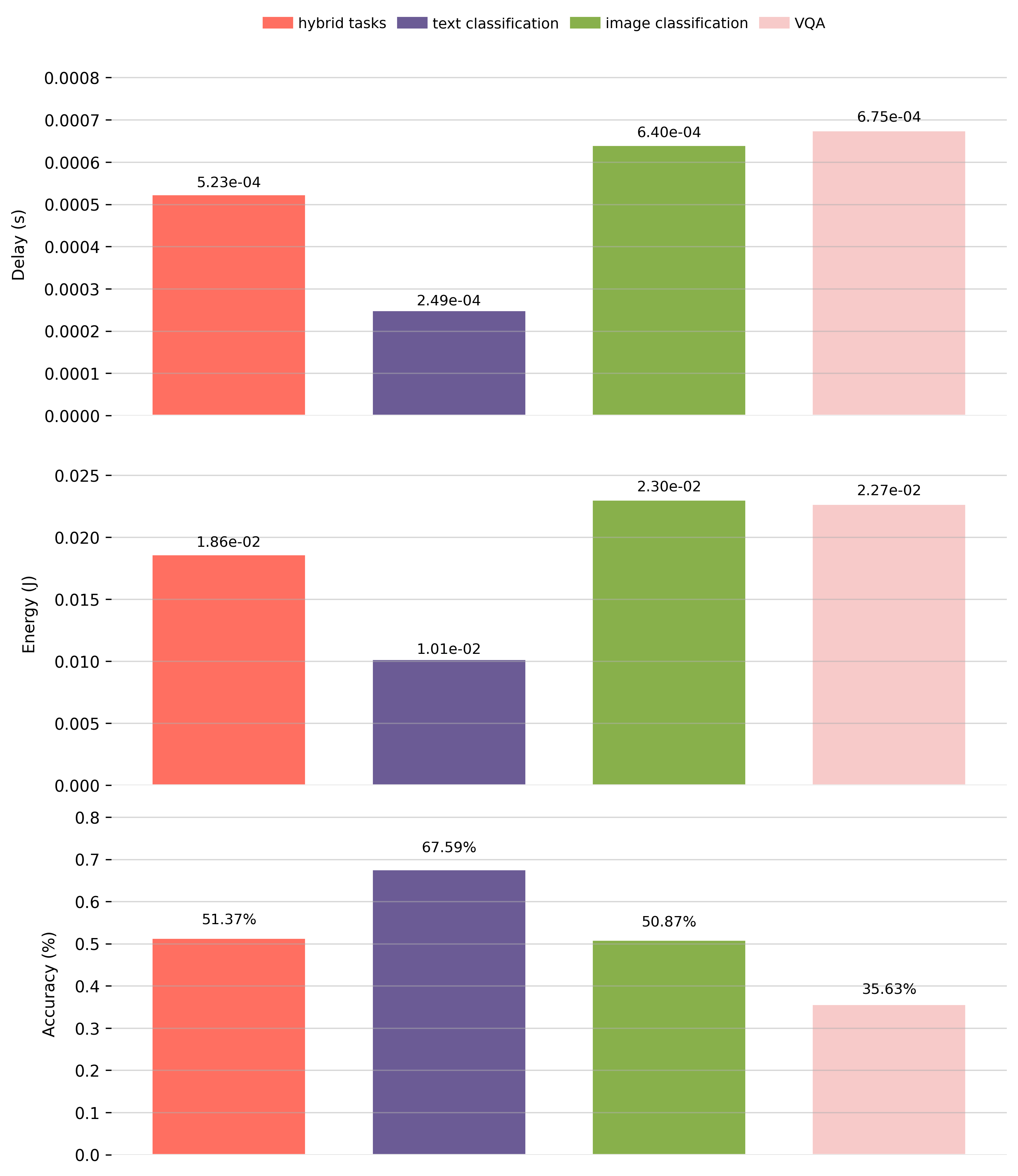}
\captionsetup{justification=raggedright}
\caption{Performance with different task types.}
\label{task_performance}
\end{figure}

Fig.~\ref{task_performance} presents the latency, energy consumption, and task performance across different task queue configurations. The system was initially trained on hybrid task scenarios and subsequently evaluated under four distinct test conditions: hybrid tasks, text classification, image classification, and VQA. The results demonstrate our algorithm's strong adaptability to diverse task types. Specifically, text classification tasks exhibit the lowest latency and energy consumption while maintaining high accuracy, attributable to their smaller data volume and computational complexity. In contrast, image classification and VQA tasks require substantially more resources, resulting in higher energy expenditure and longer processing times. The hybrid task shows performance metrics that approximately equal the arithmetic mean of individual task types, confirming the algorithm's robustness. The significant performance variations observed across task categories underscore the necessity of employing locally normalized QoE metrics for fair evaluation.

\subsection{QoE Preference}
We can adjust the model to achieve desired results by modifying QoE preferences for different metrics. Table \ref{different_preference} shows the performance in execution latency, energy consumption, and task accuracy under varying preference settings. Specifically, we adjust the relative magnitude of $\omega_t$, $\omega_e$, and $\omega_a$ to set preferences for different metrics and retrain to obtain the corresponding model for each. 1) For the delay-preference model, we set $\omega_t=1$, $\omega_e=0$, and $\omega_a=0$. The delay-preference model tends to choose a smaller $\mu_n$  to reduce the amount of data transmitted. This results in a partial loss of semantic information, causing a decline in task accuracy. 2) For the energy-preference model, we set $\omega_t=0$, $\omega_e=1$ and $\omega_a=0$. The energy-preference model prioritizes reduction of local processing energy by favoring larger semantic extraction factors, achieving significant energy savings at the cost of substantially increased transmission latency. 3) For the accuracy-preference model, we set $\omega_t=0$, $\omega_e=0$, and $\omega_a=1$. The accuracy-preference model aims to maximize the accuracy of all tasks, prompting it to execute locally. Nevertheless, Local task execution exhibits significantly higher latency and energy consumption.

\begin{figure}[t]
\centering
\includegraphics[width=0.48\textwidth]{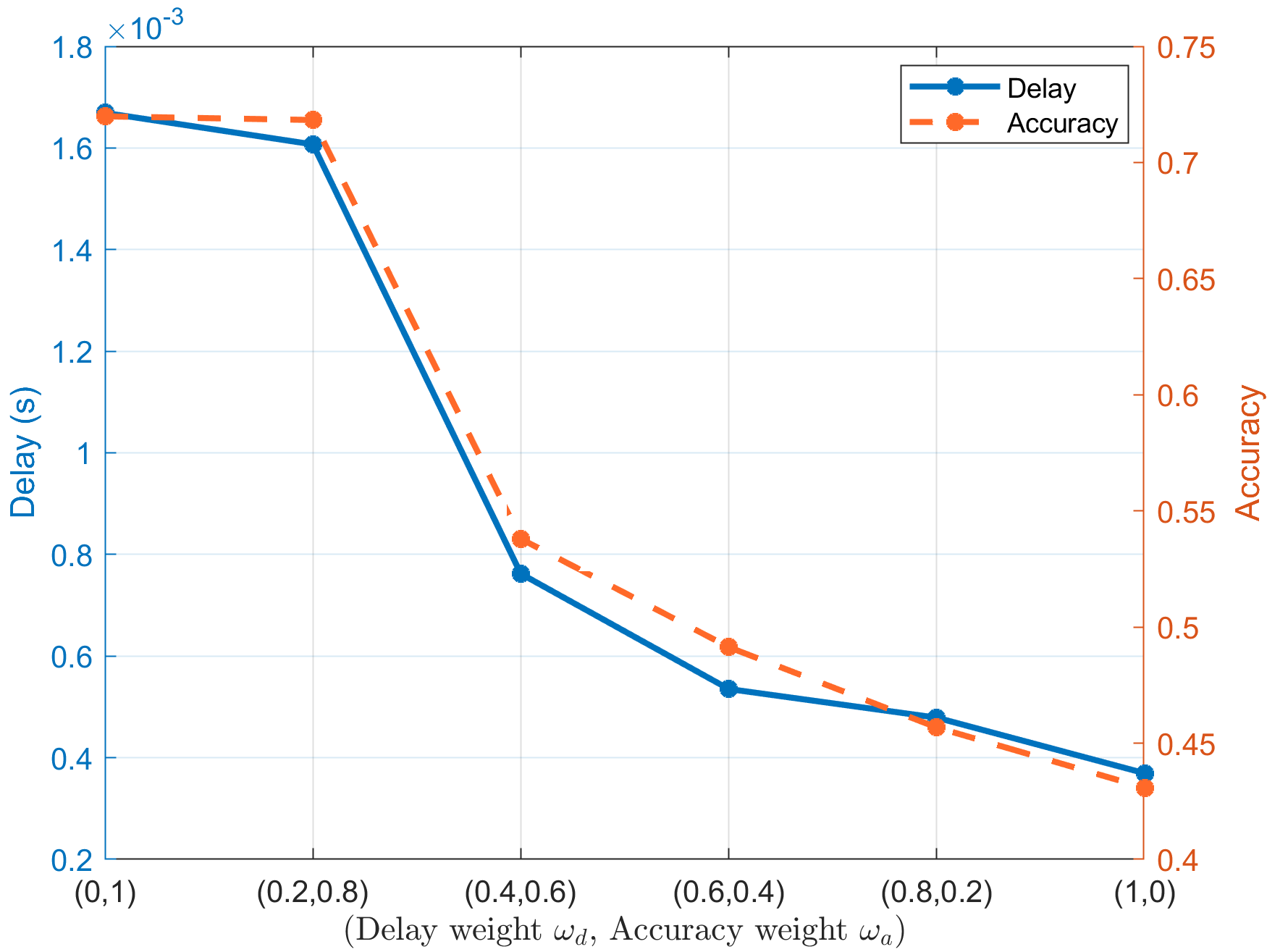}
\captionsetup{justification=raggedright}
\caption{Delay and accuracy performance with different weights.}
\label{AD-weight}
\end{figure}

\begin{figure}[t]
\centering
\includegraphics[width=0.48\textwidth]{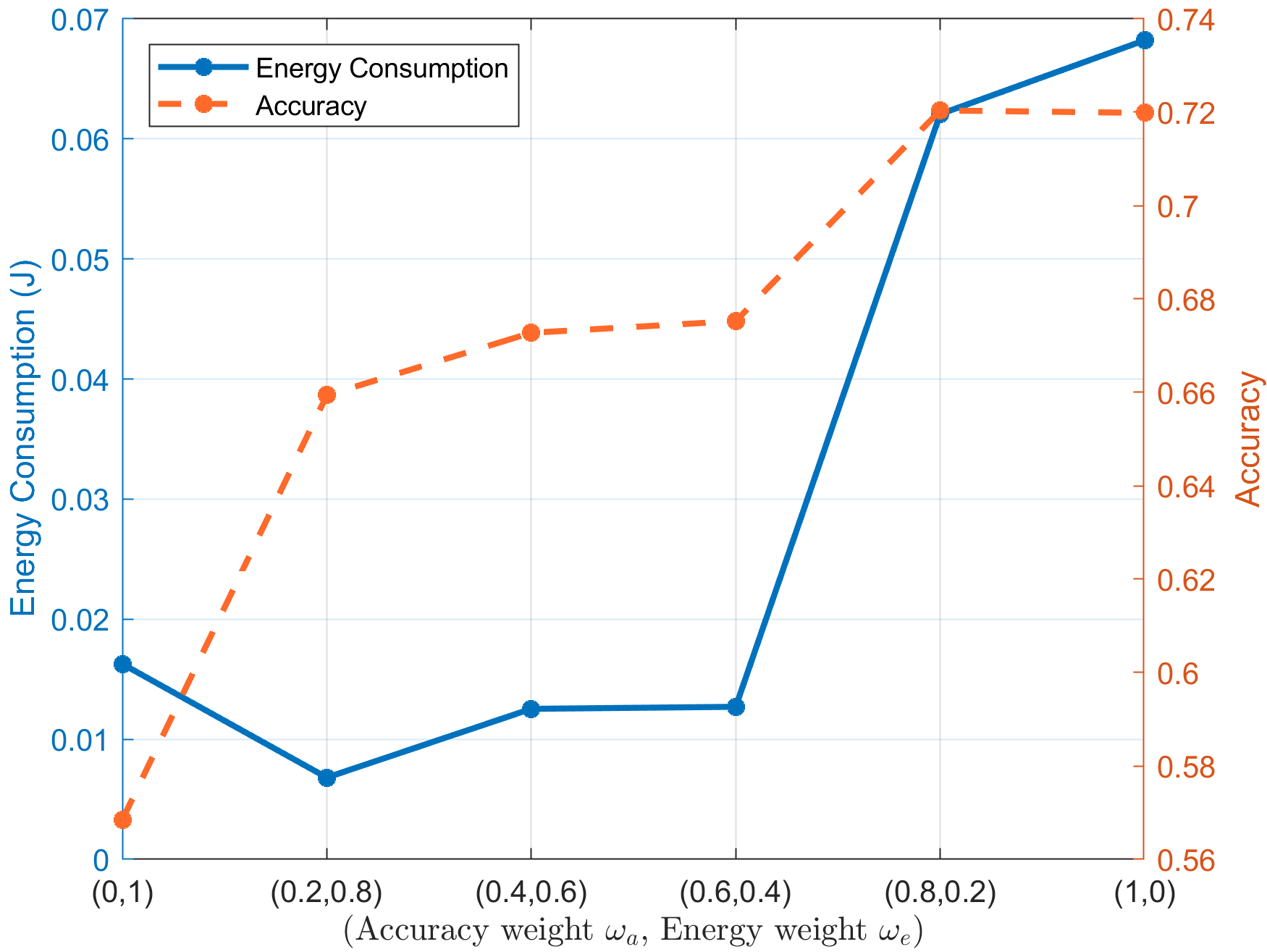}
\captionsetup{justification=raggedright}
\caption{Energy consumption and accuracy performance with different weights.}
\label{AE-weight}
\end{figure}

\begin{figure}[t]
\centering
\includegraphics[width=0.48\textwidth]{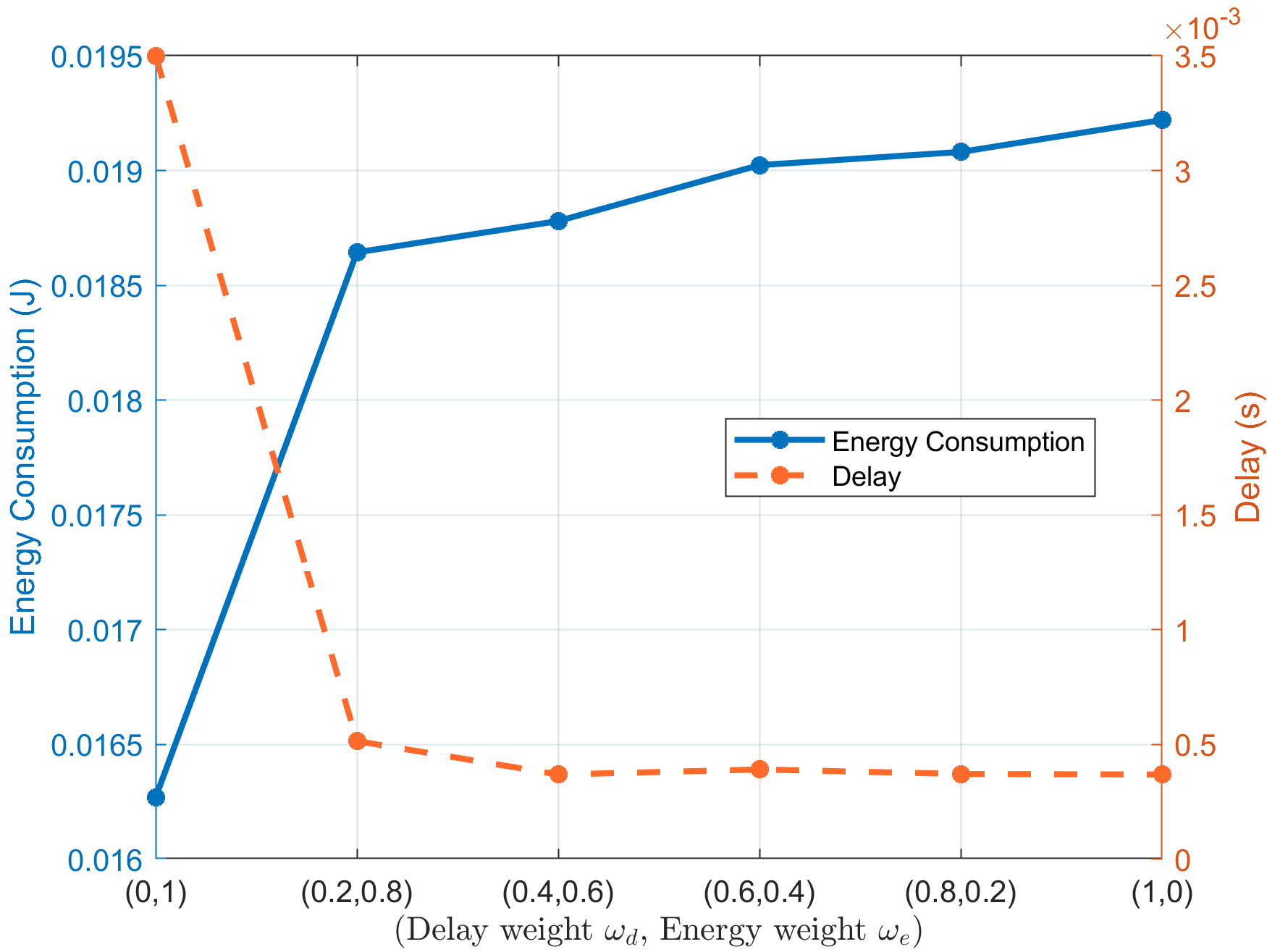}
\captionsetup{justification=raggedright}
\caption{Energy consumption and delay performance with different weights.}
\label{DE-weight}
\end{figure}

To investigate the adaptability of the proposed algorithm to different weights of latency, energy consumption, and task performance, we vary the weight settings and retrain the resource allocation model, and then evaluate the corresponding changes in latency, energy consumption, and task performance. Fig.~\ref{AD-weight} shows the performance of latency and task accuracy under different weights by adjusting the weights assigned to latency and accuracy. As the weight for latency increases, tasks are offloaded to the ES with more powerful processing capabilities, and there tends to be a preference for a smaller semantic extraction factor, $\mu_n$, to maximize the compression rate. This results in a decrease in the accuracy of tasks executed by the ES. Fig.~\ref{AE-weight} presents the performance of energy consumption and task accuracy under different weights. As the weight of task performance increases, the model allocates more tasks to local, since semantic extraction offloading degrades task performance. However, due to limited local computing capability, longer execution time is required, leading to higher energy consumption. Fig.~\ref{DE-weight} describes the performance of energy consumption and latency under varying weights. As the weight of latency increases, the model assigns more tasks to the server, leveraging its stronger computational capability, and selects a smaller semantic extraction factor to reduce transmission volume. However, extracting more compact semantic representations results in increased energy consumption.

\subsection{Sensitivity Analysis}
\begin{figure}[t]
\centering
\includegraphics[width=0.48\textwidth]{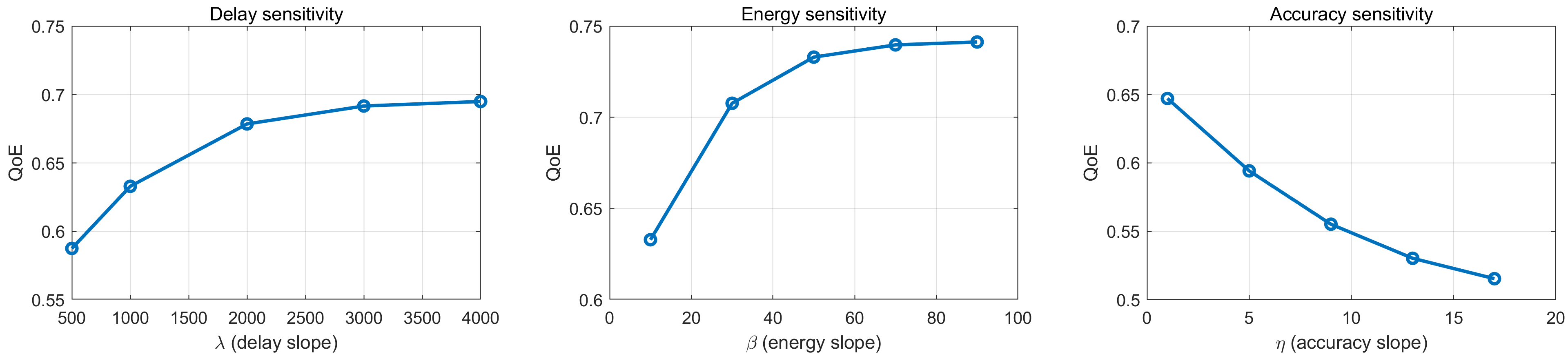}
\captionsetup{justification=raggedright}
\caption{Sensitivity analysis of logistic slope parameters.}
\label{fig:sensitivity}
\end{figure}

    We fix the trained MAPPO policy and vary one of the parameters $\lambda$, $\beta$, and $\eta$ in  \eqref{QoE} during evaluation, while keeping the other two unchanged. This allows us to isolate the impact of each parameter on the QoE mapping without introducing policy adaptation effects.
    
    Fig. \ref{fig:sensitivity} presents the sensitivity analysis results with respect to $\lambda$, $\beta$, and $\eta$. When the slope of the logistic function is excessively large (or similarly, excessively small), even a slight difference between the latency, energy consumption, or accuracy and their corresponding values under local execution will cause the logistic function to rapidly enter the saturation region. Consequently, the QoE also becomes saturated and loses sensitivity to performance variations.

    Based on this observation, we set $\lambda=10^3$, $\beta=10$, and $\eta=2$, such that the slope of the logistic function lies within an appropriate range and can effectively capture variations in latency, energy consumption, and accuracy.

\section{Conclusion}
\label{conclusion}
In this paper, we proposed a new multi-task offloading system and designed a unified quality of experience (QoE) metric to measure overall task offloading performance. Text classification, image classification, and visual question answering (VQA) tasks can be executed locally or offloaded to the edge server after semantic extraction. A semantic-aware multi-agent PPO reinforcement learning algorithm is used to solve the problem of maximizing overall user QoE through the joint optimization of the semantic extraction factor, computing resources, and communication resources. Experimental results show that the proposed method outperforms other benchmarks in overall user QoE. Specifically, the proposed algorithm demonstrates a 12.68\% improvement in QoE compared to the semantic-unaware approach. Additionally, the proposed approach can be flexibly adapted to user preferences, fully satisfying latency, energy, and task performance preference requirements. In future work, we intend to develop a robust and general semantic codec that maintains high inference accuracy across heterogeneous data modalities and stochastic, time-varying wireless channels. Besides, we plan to establish rigorous quantitative mathematical relationships between semantic extraction parameters and latency, energy consumption, and task performance, leveraging extensive empirical data.

\bibliographystyle{IEEEtran} 
\bibliography{ref}  
\begin{IEEEbiography}[{\includegraphics[width=1in,height=1.25in,clip,keepaspectratio]{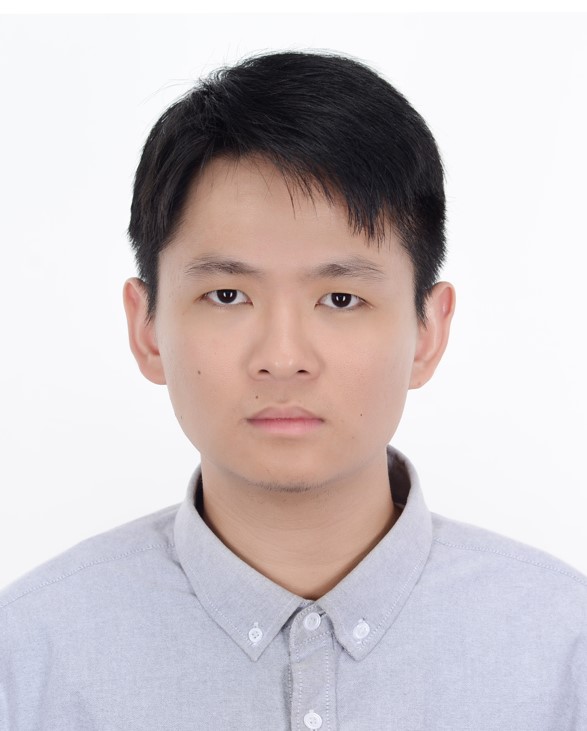}}]{Xuyang Chen}
(Student Member, IEEE) received his B.E. degree in Communication Engineering from Shenzhen University, Shenzhen, China, in 2021. He is currently pursuing a PhD in Information and Communication Engineering at Shenzhen University. His research interests include semantic communication, generative AI, resource allocation, and immersive communication, etc.\end{IEEEbiography}
\begin{IEEEbiography}[{\includegraphics[width=1in,height=1.25in,clip,keepaspectratio]{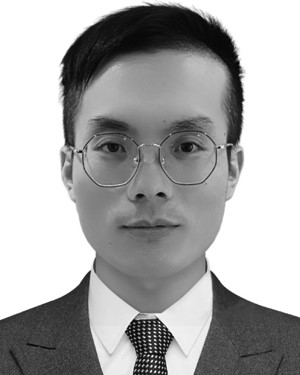}}]{Qu Luo}
(Member, IEEE) received the Ph.D. degree from the University of Surrey, U.K., in 2023. He worked at Huawei Technologies Company Ltd., Chengdu, China, from 2019 to 2020. He was also a Visiting Collaborative Researcher at the University of Essex in 2023. He is currently a Research Fellow in wireless communications at the 5GIC and 6GIC, Institute for Communication Systems, University of Surrey. His research interests include proof-of-concept physical layer design, integrated sensing and communication, non-orthogonal multiple access, random access, deep/machine learning in the physical layer, and joint MAC layer and physical layer optimization. He was a recipient of an Exemplary Reviewer Award of the IEEE Wireless Communication Letters from 2020 to 2023 and the IEEE Communication Letters in 2022 and 2023 and the Best Paper Award of the IEEE CSPS in 2018 and the IWCMC in 2024. He has also served as the workshop Chair/Co-Chair for the IEEE ICCC 25, IEEE SPAWC 25, and IEEE UCOM 25.\end{IEEEbiography}
\begin{IEEEbiography}[{\includegraphics[width=1in,height=1.25in,clip,keepaspectratio]{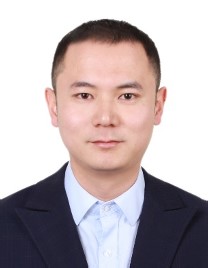}}]{Yao Sun}
 (Senior Member, IEEE) is currently a Lecturer with the James Watt School of Engineering, University of Glasgow, Glasgow, U.K. His current research interests include intelligent wireless networking, semantic communications, blockchain systems, and resource management in next-generation mobile networks. He has won the IEEE Communication Society of TAOS Best Paper Award in 2019 ICC, the 2022 IEEE INTERNET OF THINGS JOURNAL Best Paper Award, and the Best Paper Award in the 22nd ICCT.
\end{IEEEbiography}
\begin{IEEEbiography}[{\includegraphics[width=1in,height=1.25in,clip,keepaspectratio]{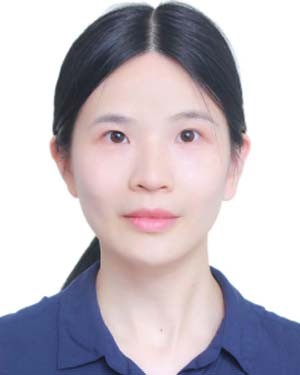}}]{Wei Jiang}
(Associate Member, IEEE) received the BS degree in School of Communication and Information Engineering from the Chongqing University of Posts and Telecommunications (CQUPT) in 2013, and the PhD degree in School of Communication and Information Engineering at University of Electronic Science and Technology of China (UESTC), in 2019. She was a visiting PhD student with Pennsylvania State University (PSU) from 2017 to 2018 and a postdoctoral Researcher with Shenzhen University from 2020 to 2022. She is currently an associate professor with the Institute of Cyberspace Security, Zhejiang University of Technology. She has won the Best Paper Award of IEEE Transactions on Services Computing in 2023. Her current research interests include next generation mobile communication systems, mobile edge computing and content caching.\end{IEEEbiography}
\begin{IEEEbiography}[{\includegraphics[width=1in,height=1.25in,clip,keepaspectratio]{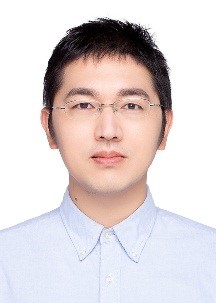}}]{Daquan Feng}
(Member, IEEE) received his Ph.D. degree in Information Engineering from the National Key Laboratory of Science and Technology on Communications, University of Electronic Science and Technology of China, Chengdu, China in 2015. He had been a visiting student in the School of Electrical and Computer Engineering, Georgia Institute of Technology, USA, from 2011 to 2014. He is currently a Distinguished Professor with the Shenzhen Key Laboratory of Digital Creative Technology, the Guangdong Province Engineering Laboratory for Digital Creative Technology, the Guangdong-Hong Kong Joint Laboratory for Big Data Imaging and Communication, College of Electronics and Information Engineering, Shenzhen University, Shenzhen, China. He has authored or co-authored over 80 journals/conferences papers, with more than 6000 citations. His research interests include edge intelligence, spatial computing, and immersive communications. Dr. Feng is an Associate Editor of IEEE Transactions on Information Forensics and Security, Science China Information Science, ICT Express and Digital Communications and Networks.
\end{IEEEbiography}
\begin{IEEEbiography}[{\includegraphics[width=1in,height=1.25in,clip,keepaspectratio]{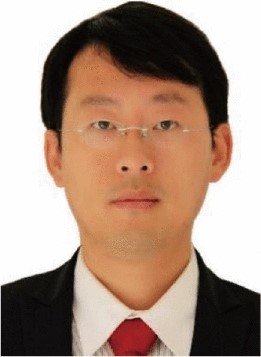}}]{Gaojie Chen}
(S’09 – M’12 – SM’18) received the B.Eng. and B.Ec. Degrees in electrical information engineering and international economics and trade from Northwest University, China, in 2006, and the M.Sc. (Hons.) and PhD degrees in electrical and electronic engineering from Loughborough University, Loughborough, U.K., in 2008 and 2012, respectively. After graduation, he took up academic and research positions at DT Mobile, Loughborough University, University of Surrey, University of Oxford and University of Leicester, U.K. He is a Professor and Associate Dean of the School of Flexible Electronics (SoFE), at Sun Yat-sen University, China, and a visiting professor at the 5GIC\&6GIC, University of Surrey, UK. His research interests include wireless communications, flexible electronics, satellite communications, the Internet of Things and secrecy communications. He received the Best Paper Awards from the IEEE IECON 2023, and the Exemplary Reviewer Awards of the IEEE WIRELESS COMMUNICATIONS LETTERS in 2018, the IEEE TRANSACTIONS ON COMMUNICATIONS in 2019 and the IEEE COMMUNICATIONS LETTERS in 2020 and 2021; and Exemplary Editor Awards of the IEEE COMMUNICATIONS LETTERS, IEEE WIRELESS COMMUNICATIONS LETTERS, and IEEE TRANSACTIONS ON COGNITIVE COMMUNICATIONS NETWORKING in 2021, 2022, 2023 and 2024, respectively. He served as an Associate Editor for the IEEE JOURNAL ON SELECTED AREAS IN COMMUNICATIONS - MACHINE LEARNING IN COMMUNICATIONS from 2021 to 2022. He serves as an Editor for the IEEE TRANSACTIONS ON WIRELESS COMMUNICATIONS, IEEE TRANSACTIONS ON COGNITIVE COMMUNICATIONS NETWORKING, IEEE WIRELESS COMMUNICATIONS LETTERS, and a Senior Editor for the IEEE COMMUNICATIONS LETTERS, and a Panel Member of the Royal Society’s International Exchanges, UK.
\end{IEEEbiography}

\end{document}